\documentclass[12pt]{iopart}

\usepackage{graphics,epsfig,subfig}
\usepackage{amssymb,amsfonts,bm}
\usepackage{mathrsfs}
\usepackage{color,psfrag}
\usepackage{iopams}
\usepackage{hhline}

\usepackage{psfrag}

\newcommand{\nn}{\nonumber}
\newcommand{\bb}{\begin{eqnarray}}
\newcommand{\ee}{\end{eqnarray}}

\renewcommand{\eref}[1]{equation (\ref{#1})}
\newcommand{\erefs}[1]{equations (\ref{#1})}
\renewcommand{\Re}{{\rm Re}}
\renewcommand{\Im}{{\rm Im}}
\renewcommand{\e}{e}

\renewcommand{\fl}{\hspace{-1cm}}

\newcommand{\vk}{v}

\newcommand{\nk}{n}
\newcommand{\ak}{A}
\newcommand{\fk}{C}
\newcommand{\dk}{F}

\newcommand{\su}{\varphi}
\newcommand{\fJ}{J}

\newcommand{\tc}{\color{black}}

\newcommand{\as}{a}
\renewcommand{\bs}{b}
\renewcommand{\equiv}{:=}
\newcommand{\parti}{d}

\begin{document}
\title[Stability condition of steady oscillations in aggregation models]{Stability condition of the steady oscillations in aggregation models with shattering process and
self-fragmentation}

\author{Jean-Yves Fortin$^1$ and MooYoung Choi$^2$}


\address{$^1$ Laboratoire de Physique et Chimie Th\'eoriques,
CNRS UMR 7019,
\\ Universit\'e de Lorraine,
F-54000 Nancy, France}

\address{$^2$ Department of Physics and Astronomy and Center for Theoretical Physics\\
Seoul National University, Seoul 08826, Korea}

\ead{jean-yves.fortin@univ-lorraine.fr, mychoi@snu.ac.kr}

\begin{abstract}{We consider a system of clusters of various sizes or masses, subject to
aggregation and fragmentation by collision with monomers or by self-disintegration. The aggregation rate for the cluster of size or mass $k$ is given by a kernel proportional to $k^{\as}$, whereas the collision and disintegration kernels are given by $\lambda k^{\bs}$ and $\mu k^{\as}$, respectively, with $0\le\as,\bs\le 1$ and positive factors $\lambda$ and $\mu$. We study the emergence of oscillations in the
phase diagram {\tc $(\mu,\lambda)$} for two models: $(\as,\bs)=(1,0)$ and $(1,1)$. It is shown that the
monomer population satisfies a class of integral equations possessing oscillatory solutions in a finite
domain in the plane {\tc $(\mu,\lambda)$}. We evaluate analytically this domain and give an
estimate of the oscillation frequency. In particular, these oscillations are found to occur generally for small but nonzero values of the parameter $\mu$, far smaller than $\lambda$.}
\end{abstract}

\section{Introduction}
%
\begin{figure}[hb]
\centering
\includegraphics[scale=0.5,angle=0,clip]{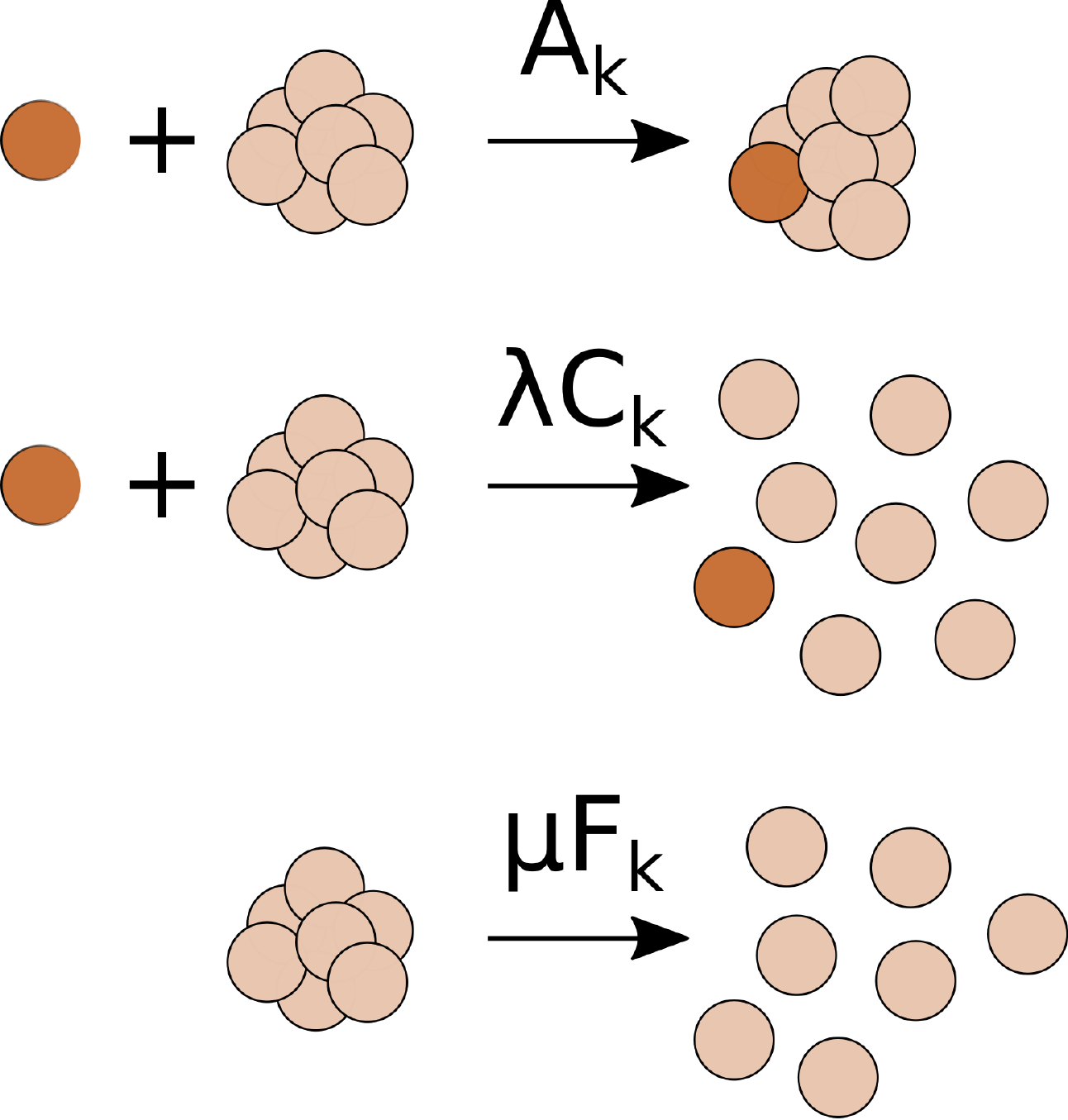}
\caption{\label{fig_model}
Aggregation, fragmentation, and disintegration processes in the model. A monomer collides and aggregates with a $k$-particle cluster with rate $\ak_k = k^{\as}$, or fragments the cluster completely with rate $\lambda\fk_k=\lambda k^{\bs}$. Self-disintegration also occurs with rate $\mu\dk_k=\mu k^{\as}$.}
\end{figure}
%
The processes of aggregation and fragmentation between clusters of particles of various sizes or masses appear ubiquitous in a large variety of physical and chemical systems. Many examples found in the literature include the coalescence of soap bubbles, stability of atmospheric particles which aggregate due to the van der Waals forces, and large-scale internet networks where nodes acquire additional links according to their attachment preference weight \cite{Dorogovtsev:2002,book:Krapivsky}. In astronomy, aggregation induced by gravitation plays an important role in forming planetary rings \cite{Cuzzi:2010} and galactic clusters \cite{Brilliantov:2015,Brilliantov:2018a}.
Fragmentation by shattering, for example, tends to counterweight aggregation by preventing the masses of the clusters from becoming too large, and this often leads to an equilibrium state of the cluster size or mass distribution which displays a power law associated with an exponential decay \cite{White:1981,Ziff:1985,Hayakawa:1987,Hayakawa:1988,Vigil:1988,Wattis:2006,Krapivsky:2017}. Fragmentation can occur by a direct collision between large clusters, and the fragmentation rate depends on various parameters, such as the cross-section or masses of the objects. This also occurs indirectly by tidal forces between large masses. Also, thermal fluctuations induce the disintegration of large polymers in solution, which plays the role of self-fragmentation.

In the long-time limit, the steady-state distribution of many models often exhibits, according to the type of kernel chosen for the aggregation or fragmentation processes, a power-law behavior associated with an exponential decay \cite{White:1981,Ziff:1985,DaCosta:2015,Krapivsky:2017,Bodrova:2019}. The exponent of the power-law decay is closely associated with the choice of the aggregation-fragmentation kernels, which depend on such parameters as the geometry of the clusters and their scattering cross-section.

{\tc Further, phase transitions can be observed while a parameter measuring the strength of the aggregation or fragmentation is varied}. For example, in the case where only monomers or one-particle clusters interact with larger ones by aggregation and fragmentation \cite{Brilliantov:2021}, if the shattering rate is too small, the monomer population will vanish, and the dynamics will stop. {\tc This phase is called jamming} and the population of {\tc the larger} clusters will therefore depend on the initial conditions. Otherwise, the distribution will {\tc relax towards} a steady-state {\tc independent of the initial conditions}. In between, {\tc for a special value of this parameter}, there exists a critical phase where only giant clusters with large masses are created and where the population of other smaller ones tends to vanish asymptotically.
{\tc Steady-state oscillations in cluster size densities can also be observed when a source of monomers and seed-clusters are introduced in an open system, with oscillation amplitude decreasing as the cluster size increases \cite{Matveev:2020}. This phenomenon contrasts from the usual view of the cluster density decreasing uniformly with the size.}

Besides time-independent equilibrium distributions of masses, one also observes collective and stable {\tc time} oscillations in a narrow window of parameter range \cite{Ball:2012,Matveev:2017,Connaughton:2018,Pego:2020,Slomka:2020,Budzinskiy:2021,Kalinov:2022,Niethammer:2022,Fortin:2022}.
This arises from {\tc nonlinear} effects or imbalances between aggregation and fragmentation processes, with or without an external source of particles supplied to the system, {\tc with a competition between time scales.}

In a recent publication \cite{Budzinskiy:2021}, oscillations were observed in a model where aggregation between monomers and clusters of size $k$ occurs with rate $A_k=k$, as well as self-disintegration
process with rate $\mu k^{\beta}$, {\tc $\beta>1$}. They observed oscillations in a domain where $\mu$ is smaller than $10^{-6}$ and exponent $\beta$ smaller than $6$ after linearizing the dynamical equations around the constant solution, reducing the problem to a linear algebra system. The {\tc existence of oscillations is investigated by evaluating} the complex eigenvalues for a finite set of cluster sizes, and the study of the sign of the real part of the eigenvalues close to the origin as well as the imaginary part characterizes the {\tc stability} of the perturbation. The transition {\tc from stable and constant or steady-state solution} to oscillatory solutions is often called Hopf bifurcation. {\tc In the present manuscript
we do not solve for the Jacobi matrix associated to the dynamical equations but rather present another method and show that the monomer density satisfies an exact integral equation containing oscillatory solutions, without approximation and for an infinite set of cluster sizes. A perturbation method then
allows us to check the stability of the steady-state solution.}

Usually, aggregation and fragmentation dynamics are described through Smoluchowski equations, which govern the fluctuations of the cluster masses. These Smoluchowski equations are quadratic in densities, and exact solutions may not be obtained in general, except for small sets of variables \cite{Buchstaber:2012a, Calogero:2020} in the mathematical field of nonlinear systems of quadratic equations.
In this case, it is difficult to characterize the presence of density oscillations. However, if clusters interact only with monomers, the dynamical equations can be simplified by rescaling all the densities and redefining time. All the properties of the dynamics can be deduced from an integral equation involving solely the distribution density of the monomers. This allows us to study analytically the oscillation behavior of the system in more detail. In case a source of the monomer is present \cite{Fortin:2022}, within some approximation, the monomer density satisfies a nonlinear ordinary differential equation (ODE) with a fluctuating damping term, {\tc or time-dependent factor in front of the first time derivative of the density}. This damping term is partially responsible for the Li\'enard oscillations that are observed {\tc in this model}: When the damping term is positive, the monomer density reduces toward zero. However, as the amplitude of the density gets very small, the damping term changes sign and forces the density to grow again. {\tc We expect that} the presence of a source or even a self-fragmentation process constraints the monomer density to remain positive.

In this paper, we investigate a model restricted to dynamics driven by monomers, which is considered an extension of the Becker-D\"oring model \cite{Pego:2020,Niethammer:2022} {\tc involving reactions with monomers only}. The dynamics includes not only aggregation between monomers and larger clusters as well as fragmentation or shattering of clusters by collision with monomers but also self-fragmentation of clusters into monomers when they become too large. {\tc The presence of a self-disintegration process allows the system to have a continuous source of monomers since large clusters are always present. In the recent reference \cite{Niethammer:2022},
the Becker-D\"oring model is modified by injecting monomers (source) and depleting the largest clusters (drain). This model presents oscillations due physically to desynchronization between the source and removal mechanism.} The presence of self-fragmentation prevents the population of monomers from vanishing completely and the dynamics from stopping in a system where mass conservation holds. {\tc A source of monomers would require a mechanism to remove clusters in order to preserve the total mass. Our model is an extension of the model studied in reference \cite{Brilliantov:2021}, where no intrinsic oscillations are observed, as self-disintegration process is incorporated. We will see that the combination of aggregation, fragmentation and self-disintegration leads to steady time oscillations, which originates from a complex mechanism between the three processes, as each of these possesses a different time scale. We investigate in this paper the domain of parameters for which oscillations are the stable state, unlike the time-independent steady state.} We would like particularly to develop a technical method applicable to the study of collective oscillations through the resolution of differential equations satisfied by a class of generating functions. These generating functions contain all the moments of the densities, and in particular, we can extract precisely the monomer density, which determines all the other densities in a closed form. Computing eigenvalues in the analysis of the stability of the oscillations is replaced by finding zeros of special functions in the complex plane for an infinite system.
{\tc We will generally call an oscillatory solution stable if any small perturbation of the steady-state or time-independent density of the monomers for example, which is also a solution of the dynamical system asymptotically and independently of the initial conditions, grows with time and presents an oscillating behavior with a finite frequency.}

{\tc To summarize the results, we found the integral equations for the monomer density in two models with different aggregation, self-fragmentation, and fragmentation rates, using the characteristics method to solve the generating function. These integral equations possess explicit oscillatory solutions. Using a perturbation theory, we obtain the exact phase diagram of the existence of these oscillations as function of the parameters describing the fragmentation and self-fragmentation rates. The precise tuning and balance between the two processes of fragmentation lead to oscillations which are persistent in the long time limit.}

The paper is organized as follows: Section \ref{sec_model} introduces the main model with the general dynamics, from which two cases with different exponents are studied in detail. 
We present a method based on the differential equation satisfied by a generating function and display the phase diagram of the domain of oscillation stability as a function of the parameter amplitudes. In section \ref{sec_disc}, we discuss the general class of the obtained generating functions and integral equations for which oscillating solutions exist. Finally, section \ref{sec_concl} gives a brief conclusion of the present study.

\section{Model \label{sec_model}}

We consider a set of clusters made of $k$ particles, each with a mass probability density $n_k$. Only
one-particle clusters react with bigger clusters of size $k\ge 2$. A particle can
aggregate with a cluster of size $k$ to form a cluster of size $k+1$ with rate $\ak_k$.
Or it can fragment the cluster completely into $k$ single particles with rate $\lambda\fk_k$.
{\tc We assume that the aggregation rate $\ak_k$ grows with the cluster size $k$, according to a power law with exponent $\as$, such that $\ak_k = k^{\as}$. Similarly, the fragmentation rate is chosen such that $\lambda\fk_k=\lambda k^{\bs}$ with $\bs\le\as$. Clusters of size $k$ can spontaneously disintegrate into $k$ monomers with a rate chosen to be proportional to the aggregation rate $\mu\dk_k =\mu k^{\as}$, with a coefficient of proportionality. We indeed assume that the disintegration probability should increase with the size of the cluster: bigger clusters tend to self-fragment more easily than smaller clusters. We could take a disintegration kernel with exponent $c$, but for simplicity here we take $a=c$. The coefficient $\mu$, generally small, reflects the fact that the process does not occur too often.}
All these processes are illustrated in figure \ref{fig_model}, {\tc and the kernels satisfy homogeneity property as $\fk_{sk}=s^{\as}\fk_k$ and $\dk_{sk}=s^{\bs}\dk_k$}. Normally the aggregation rate is proportional to the cluster size or surface, eventually fractal, and it is usual to take $0\le\as\le 1$, as observed for the problem of network growth \cite{Krapivsky:2001,Dorogovtsev:2002}. {\tc If $\as=1$, the rate of aggregation is proportional to the cluster size and this process is called linear. In the case $\as<1$,
the attachment of monomers to clusters is sublinear. When $\as=0$, the aggregation rate is independent of the size. However when $\as>1$, the attachment to the largest clusters is favored and the process is superlinear \cite{Krapivsky:2001}.
For the fragmentation exponent $\bs$, it will also be taken less than or equal to unity.}

{\tc We consider a system made of clusters of different sizes, with $N$ denoting the size of the largest cluster. $N$ is a fixed parameter independent of time that will be finite only for the numerical analysis and taken to infinity in analytical formulas. Taking into account all the processes, we have the following master equations for the densities:}
\bb\nn
\frac{\parti \nk_1}{\parti t}=
-2\nk_1^2-\sum_{k=2}^{N-1}(\ak_k-\lambda k\fk_k)\nk_1\nk_k+\lambda N\fk_N\nk_1\nk_N
+\sum_{k=2}^{N}\mu k\dk_k\nk_k,
\\ \nn
\frac{\parti \nk_k}{\parti t}=
\ak_{k-1}\nk_1\nk_{k-1}-(\ak_k+\lambda\fk_k)\nk_1\nk_k-\mu\dk_k\nk_k  \;~~~ (2\le k\le N-1),
\\ \label{eq_gen_nk}
\frac{\parti \nk_N}{\parti t}=
\ak_{N-1}\nk_1\nk_{N-1}-\lambda\fk_N\nk_1\nk_N-\mu\dk_N\nk_N .
\ee
%
%
{\tc In order to close the equations, we impose that the largest cluster does not aggregate with
monomers, so that no new larger clusters are created.} These coupled equations satisfy the conservation of the total mass, set equal to unity: $\sum_{k=1}^Nk\nk_k=1$ or $\parti(\sum_{k=1}^Nk\nk_k)/\parti t=0$.
It is convenient to redefine the variables by introducing $\vk_k \equiv \ak_k\nk_k$
and effective time $\tau\equiv\tau(t)$ such that $d\tau = \nk_1 (t)dt$.
This leads to a set of quasi-linear ODEs:
\bb\nn
&\frac{\parti \vk_1}{\parti\tau}=-2\vk_1-\sum_{k=2}^{N-1}(1-\lambda k^{1+\bs-\as})\vk_k
+\lambda N^{1+\bs-\as}\vk_N+\sum_{k=2}^{N}\mu k \frac{\vk_k}{\vk_1},
\\ \nn
&\frac{\parti \vk_k}{\parti \tau}=
k^a\vk_{k-1}-(k^{\as}+\lambda k^{\bs})\vk_k-\mu k^{\as} \frac{\vk_k}{\vk_1}  \;~~~ (2\le k\le N-1),
\\ \label{eq_gen_vk}
&\frac{\parti \vk_N}{\parti \tau}=
N^{\as}\vk_{N-1}-\lambda N^{\bs}\vk_N-\mu N^{\as} \frac{\vk_N}{\vk_1} .
\ee
{\tc In the following sections, \ref{sec_10} and \ref{sec_11}, we will study two cases of interest: $(\as,\bs)=(1,0)$ and
$(\as,\bs)=(1,1)$, as they correspond to typical cases where the aggregation rate is proportional
to the target cluster size, and the fragmentation rate is either proportional ($\bs=1$) or independent ($\bs=0$) of the cluster size.}
\subsection{Model with $(\as,\bs)=(1,0)$ \label{sec_10}}
In this section, we choose the parameters $(\as,\bs)=(1,0)$ and take the limit
$N\rightarrow\infty$. Then the master equations read
\bb\nn
&\frac{\parti \vk_1}{\parti\tau}=-2\vk_1-\sum_{k\geq 2}(1-\lambda)\vk_k
+\sum_{k\geq 2}\mu k \frac{\vk_k}{\vk_1},
\\ \label{eq_vk_1}
&\frac{\parti \vk_k}{\parti \tau}=
k\vk_{k-1}-(k+\lambda)\vk_k-\mu k \frac{\vk_k}{\vk_1}   \;~~~(k\ge 2).
\ee
It is useful to define the set of moments: $\su_l \equiv \sum_{k\geq 2}k^l\vk_k$.
Since the total mass is given by $\sum_{k=1}^N k\nk_k= \sum_{k=1}^N v_k =1$, we have 
the identity $\su_0=1-\vk_1$, and we can write the first equation of the system (\ref{eq_vk_1}) in the form:
$\partial\vk_1/\partial\tau=\lambda-1 - (1+\lambda)\vk_1+\mu\su_1/\vk_1$.
The general system of ODEs associated to $\su_l$ is given by
\bb\label{eq_phi1}
\frac{\parti \su_l}{\parti\tau}=-\left(1+\frac{\mu}{\vk_1}\right)\su_{l+1}-\lambda\su_l+2^{l+1}\vk_1
+\sum_{j=0}^{l+1}{l+1\choose j}\su_j,
\ee
for $l\ge 0$. We then introduce a generating function $G(u,\tau)\equiv\sum_{l\ge 0}\su_lu^l/l!$,
which satisfies the partial differential equation (PDE)
\bb\label{eq_G}
\frac{\partial G(u,\tau)}{\partial\tau}+\left(1+\frac{\mu}{\vk_1}-e^u\right)\frac{\partial G(u,\tau)}{\partial u}
=(e^u-\lambda)G(u,\tau)+2e^{2u}\vk_1,
\ee
with the initial condition $G(u,0)$ depending on the initial cluster densities.
Note that in the derivation of \eref{eq_G}, we have used the formula
\bb\label{eq_formula}
\sum_{l\ge \max(0,j-1)}{l+1\choose j}\frac{u^{l}}{l!}=\frac{u^{j-1}(u+j)}{j!}e^u .
\ee
This can be demonstrated using the Egorychev method and the representation of the binomial
coefficient in terms of a complex integral with a closed contour around the origin:
\bb\label{eq_binom}
{l+1\choose j}=\oint \frac{dz}{2i\pi z}\frac{(1+z)^{l+1}}{z^j},
\ee
which reduces the summation over $l$ in \eref{eq_formula} to a simple complex integral evaluated as follows:
\bb
\sum_{l\ge 0}{l+1\choose j}\frac{u^{l}}{l!}
=\oint \frac{dz}{2i\pi z}\frac{(1+z)}{z^j}e^{u(1+z)}=\frac{u^{j-1}(u+j)}{j!}e^u .
\ee
Now, {\tc \eref{eq_G} is a linear differential equation which relates implicitly $G$ as a function of $\vk_1$. We can obtain a closed form for $\vk_1$ by noticing that $G(0,\tau)=\su_0=\sum_{k\ge 2}\vk_k=1-\vk_1$. In this context, we have eliminated all dependence on the other cluster densities and isolate
a relation involving only $\vk_1$.
The following analysis consists in solving the PDE for $G$, and this is possible
as it satisfies a linear differential equation with first order derivatives.}
In order to solve \eref{eq_G}, we apply the Lagrange-Charpit method based on the characteristic curves
\cite{book:Copson,book:Polyanin2003}.
We first parametrize $u$ and $\tau$ by an external variable $s$, such that $\left(u(s),\tau(s)\right)$ defines
a curve on the surface $\left(u,\tau,G(u,\tau)\right)$. The generating function becomes a function of $s$, which we denote $\tilde G(s)\equiv G(u(s),\tau(s))$. The derivative $\tilde G'(s)$ is then given, via the chain rule, by $\tilde G'(s)=u'(s)\partial_u G(u(s),\tau(s))+\tau'(s)\partial_{\tau}G(u(s),\tau(s))$.
For convenience, we choose the variable $s$ such that
\bb
\tau'(s)=1 \; \mbox{and} \; u'(s)=1+\frac{\mu}{\vk_1(\tau(s))}-e^{u(s)}.
\ee
These can be integrated into
\bb\nn
\tau(s)=s, \\
u(s)=\int_0^s ds' \left(1+\frac{\mu}{\vk_1(s')}\right)
-\log\left(C+\int_0^s ds' \exp\left [\int_0^{s'}ds'' \left(1+\frac{\mu}{\vk_1(s'')}\right)\right ]\right),
\ee
where $C$ is a constant determined by the boundary conditions on the curve.
Imposing here that the point $(u,\tau)$ belongs to the curve, i.e., $s=\tau$ and $u(\tau)=u$, we obtain
the constant value
\bb
C=\exp\left (\gamma(\tau)-u\right )-\int_0^{\tau}ds
\exp\left (\gamma(s)\right ),
\ee
%
where $\gamma(s)\equiv s+\int_0^{s} ds' \mu\vk_1^{-1}(s')$.
With such parametrization, \eref{eq_G} becomes a first-order ODE
\bb\label{eq_Gs}
\tilde G'(s)
=(e^{u(s)}-\lambda)\tilde G(s)+2e^{2u(s)}\vk_1(s) ,
\ee
which can be integrated into
\bb\nn
\tilde G(s)=2e^{W(s)}\int_0^s ds' \vk_1(s')e^{-W(s')+2u(s')},
\\
W(s)=\int_0^s ds' e^{u(s')}-\lambda s = \log\left(C+\int_0^s ds' e^{\gamma(s')}\right) -\log C -\lambda s .
\ee
After simplifying the previous expressions, we obtain $\tilde G(s)$ in the form
\bb\nn
\tilde G(s)&=&2
\left( e^{-u}-
\int_{s}^{\tau} ds''\exp\left[s''-\tau-\mu\int_{s''}^{\tau} ds''' \vk_1^{-1}(s''')\right]\right)
\\ & & \times
\int_0^s ds' \vk_1(s')\frac{\exp\left[
\lambda(s'-s)+2(s'-\tau)-2\mu\int_{s'}^{\tau}ds''\vk_1^{-1}(s'')\right]}{\left(
e^{-u}-\int_{s'}^{\tau}d\tau'' \exp\left[s''-\tau-\mu\int_{s''}^{\tau} ds''' \vk_1^{-1}(s''')\right]
\right)^3} .
\ee
The initial conditions are chosen such that only monomers are present at
$s=\tau=0$ with the total mass unity, namely, $\vk_k(0)=\delta_{k1}$. This implies that $\tilde G(0)=G(u(0),0)=0$ since $\su_l$ is initially zero for all $l$.
Substituting $s=\tau$ and $u=0$, we obtain an integral equation for $\vk_1(\tau)$. Indeed, {\tc we have $G(0,\tau)=1-\vk_1$, which leads to the formula}
\bb\label{eq_v1}
\vk_1(\tau)=1-2
\int_0^{\tau}d\tau' \vk_1(\tau')\frac{\exp\left[
-(2+\lambda)(\tau-\tau')-2\mu\int_{\tau'}^{\tau}d\tau'' \vk_1^{-1}(\tau'')\right]}{\left(
1-\int_{\tau'}^{\tau}d\tau'' \exp\left[\tau''-\tau-\mu\int_{\tau''}^{\tau}d\tau''' \vk_1^{-1}(\tau''')\right ]
\right)^3} .
\ee
\subsubsection{Case $\mu=0$}
When $\mu=0$, \eref{eq_v1} can be simplified and reduces to
\bb\label{eq_v1_mu0}
\vk_1(\tau)=1-2\int_0^{\tau}d\tau' \vk_1(\tau')e^{-(\lambda-1)(\tau-\tau')} .
\ee
Taking the Laplace transform, we obtain, for $\lambda>1$,
\bb
\hat \vk_1(p)=\frac{1}{p}\frac{\lambda-1+p}{\lambda+1+p},
\ee
which gives a long-time {\tc steady-state, or constant value since $\vk_1$ reaches a constant in the long time limit,} $\vk_1(\tau)\simeq (\lambda-1)/(\lambda+1)$ as $p\rightarrow 0$.
We can also differentiate \eref{eq_v1_mu0} and eliminate the integral to obtain the ODE
\bb
\vk_1'(\tau)=-(1+\lambda)\vk_1(\tau)+\lambda-1 ,
\ee
which yields the time-dependent solution $\vk_1(\tau)=(\lambda-1+2e^{-(\lambda+1)\tau})/(\lambda+1)$.
For $\lambda<1$, $\vk_1$ decreases and vanishes at a finite time
$\tau=(1+\lambda)^{-1}\log(2/[1-\lambda])$ {\tc where the dynamics stops. Otherwise, $\vk_1$ decreases and reaches the constant solution $(\lambda-1)/(\lambda+1)$ after a transitory regime.}
\subsubsection{Steady-state solution}
\begin{figure*}[ht]
\centering
\includegraphics[angle=0,scale=0.5,clip]{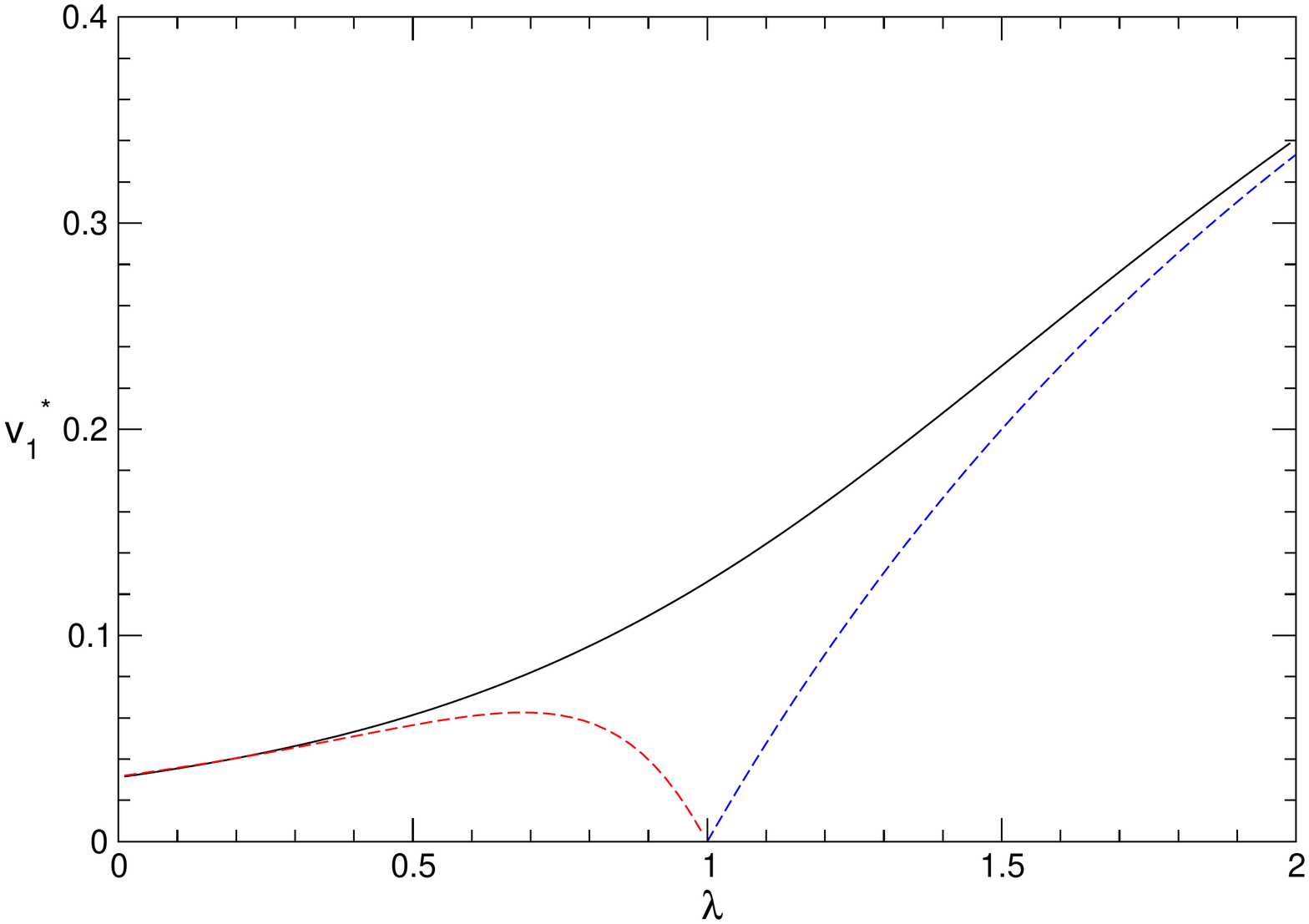}
\caption{\label{fig_sol_v1}
{\tc Solution $\vk_1^*$ given by solving \eref{eq_v1eq2} for $\mu=10^{-3}$ (black line). 
The dashed blue line is the result
for $\mu=0$, or \eref{eq_v1_lambda_large}. The dashed red line is the approximation for $\lambda<1$
and given by \eref{eq_v1_lim}}.}
\end{figure*}

Here we discuss {\tc the existence of a long-time constant solution of \eref{eq_v1} when $\mu$ is non zero}.
Supposing that $\vk_1(\tau)$ reaches a limit $\vk_1^* \,(>0)$, we obtain the equation describing this solution when $\tau\gg 1$:
\bb\nn
\vk_1^* &\simeq& 1-2\vk_1^*
\int_0^{\tau} d\tau' \frac{e^{
-(2+\lambda+2\mu/\vk_1^*)(\tau-\tau')}}{\left(
1-\int_{\tau'}^{\tau}e^{-(1+\mu/\vk_1^{*})(\tau-\tau'')}d\tau''
\right)^3}
\\ \label{eq_v1eq}
&\simeq& 1-2\vk_1^*(1+\epsilon)^3\int_0^{\tau}d\tau' \frac{e^{-(2+\lambda+2\epsilon)\tau'}}{
\left(\epsilon+e^{-(1+\epsilon)\tau'}\right)^3} ,
\ee
where $\epsilon \equiv\mu/\vk_1^*$ is assumed small ($\epsilon\ll 1$) for small $\mu$.
When $\mu=0$ or $\epsilon=0$ and $\lambda>1$, the integral in \eref{eq_v1eq} is finite, and $\vk_1$ approaches exponentially to the constant solution $\vk_1^*=(\lambda-1)/(\lambda+1)$
as $\tau\rightarrow\infty$, as seen in the previous subsection. Otherwise, for $\lambda<1$ ($\mu=0$), $\vk_1$ reaches zero at some finite time
$\tau$ and the dynamics stops \cite{Brilliantov:2021}. {\tc For the special value $\lambda=1$, $\vk_1$ decays to zero exponentially as $\vk_1(\tau)=\e^{-2\tau}$, while
$n_1(t)$ decays to zero with a power law as $n_1(t)=1/(2t+1)$, after the original time variable was replaced}. When $\mu\neq 0$, there is a finite nonzero solution to the integral \eref{eq_v1eq} for any value of $\lambda$.
Changing the variable $x=\epsilon^{-1}e^{-(1+\epsilon)\tau'}$ and taking the limit $\tau\rightarrow\infty$, we rewrite \eref{eq_v1eq} as
\bb\label{eq_v1eq2}
\vk_1^*=\left[
1+2(1+\epsilon)^2\epsilon^{\frac{\lambda-1-\epsilon}{1+\epsilon}}
\int_0^{1/\epsilon} dx \frac{x^{\frac{1+\lambda+\epsilon}{1+\epsilon}}}{(1+x)^3}
 \right]^{-1}.
\ee
The behavior of the integral depends on $\lambda$ when $\epsilon$ is small.
If $\lambda>1+\epsilon$, the integral diverges for large $x$, and the dominant contribution
comes from $x\simeq 1/\epsilon\,(\gg 1)$:
\bb
\int_0^{1/\epsilon} dx \frac{x^{\frac{1+\lambda+\epsilon}{1+\epsilon}}}{(1+x)^3}
 \simeq \frac{1+\epsilon}{(\lambda-1-\epsilon)}\epsilon^{\frac{1+\epsilon-\lambda}{1+\epsilon}}.
\ee
Combining this asymptotic result with \eref{eq_v1eq2}, we obtain, for $\epsilon\ll 1$:
\bb\label{eq_v1_lambda_large}
{\vk_1^*}=\left[1+2\frac{(1+\epsilon)^3}{(\lambda-1-\epsilon)}
\right]^{-1}\simeq \frac{\lambda-1}{\lambda+1} ,
\ee
which is the expected result for $\lambda$ large compared with $\mu$. If
$\lambda<1+\epsilon$, then the integral converges in the limit $x\rightarrow\infty$:
\bb
\int_0^{\infty} dx \frac{x^{\frac{1+\lambda+\epsilon}{1+\epsilon}}}{(1+x)^3} =
\frac{\pi\lambda(\lambda+1+\epsilon)}{2(1+\epsilon)^2\sin(\pi\lambda/(1+\epsilon))},
\ee
which is a finite quantity. Therefore the dominant term in \eref{eq_v1eq2} is $\epsilon^{\frac{\lambda-1-\epsilon}{1+\epsilon}} \,(\gg 1)$, and some algebra leads to the approximate expression:
\bb\label{eq_v1_lim}
\vk_1^*\simeq \mu^{(1-\lambda)/(2-\lambda)}\left[\frac{\sin(\pi\lambda)}{\pi\lambda(\lambda+1)}
\right]^{1/(2-\lambda)},
\ee
for $\lambda<1$ and $\mu\ll 1$.
It is then straightforward to verify that $\epsilon=\mu/\vk_1^*\propto \mu^{1/(2-\lambda)}\ll 1$.
{\tc The quantity (\ref{eq_v1_lim}) is positive and non zero for any small value of $\mu$.
In figure \ref{fig_sol_v1} we have represented the exact solution of \eref{eq_v1eq2} as function of 
$\lambda$ for a given small value of $\mu=10^{-3}$, as an illustration purpose (black line). It is compared with the solution at $\mu=0$, or \eref{eq_v1_lambda_large}. The two solutions coincide asymptotically for $\lambda
\gg 1$. In the opposite case, for $\lambda<1$, the solution is well represented by \eref{eq_v1_lim}
as $\lambda\rightarrow 0$ or $\vk_1^*$ small. For the special case $\lambda=1$, it is worth noticing that \eref{eq_v1_lim} vanishes.}

%
{\tc In the steady-state regime, the other densities $\vk_k^*$ are proportional to $\vk_1^*$
and can be deduced from the steady-state solution of \eref{eq_vk_1}
\bb
\vk_k^*=\frac{\Gamma(k+1)\Gamma(2+\lambda/(1+\epsilon))}{\Gamma(k+1+\lambda/(1+\epsilon))
(1+\epsilon)^{k-1}}\vk_1^*,
\ee
which asymptotically decreases exponentially like $\vk_k^*\propto k^{-\lambda/(1+\epsilon)}e^{-k\log(1+\epsilon)}$, with a power law factor whose exponent depends on $\vk_1^*$ through the parameter $\epsilon$. We notice that when 
$\epsilon\ll 1$, all the $\vk_k$s almost decrease like a power law $\vk_k\propto k^{-\lambda}$,
or $n_k\propto k^{-\lambda-1}$. Also, the conservation condition $\sum_{k\ge 1}\vk_k=1$ gives an implicit expression for $\vk_1^*$ but in term of a hypergeometric 
function
\bb
{}_2F_1\left (1,2;2{+}\frac{\lambda}{1{+}\epsilon};\frac{1}{1{+}\epsilon}\right )\vk_1^*=1.
\ee
}
We can check numerically that this expression is equivalent to the integral \eref{eq_v1eq2}.
\subsubsection{Stability around the {\tc steady-state} solution}
\begin{figure*}[ht]
\centering
\includegraphics[angle=0,scale=0.6,clip]{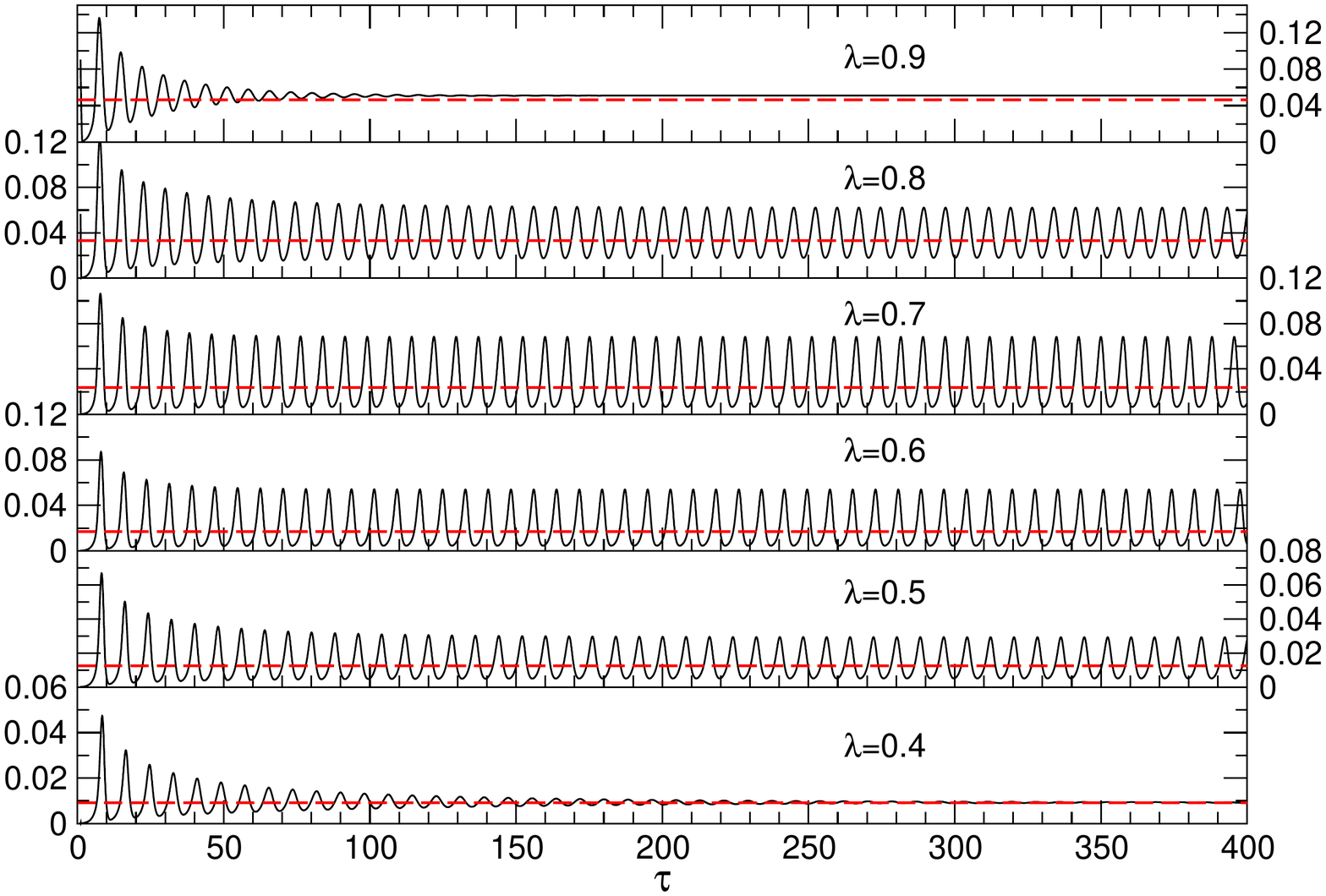}
\caption{\label{fig_oscN1000}
Time evolution of $\vk_1(\tau)$ for $\mu=10^{-5}$, $N=1000$, and several values of $\lambda$. For $\lambda\approx 0.6$, one observes steady oscillations around the {\tc steady-state} solution $\vk_1^*$ (dashed lines) given by \eref{eq_v1eq2}.}
\end{figure*}

\begin{figure*}[ht]
\centering
\includegraphics[angle=0,scale=0.6,clip]{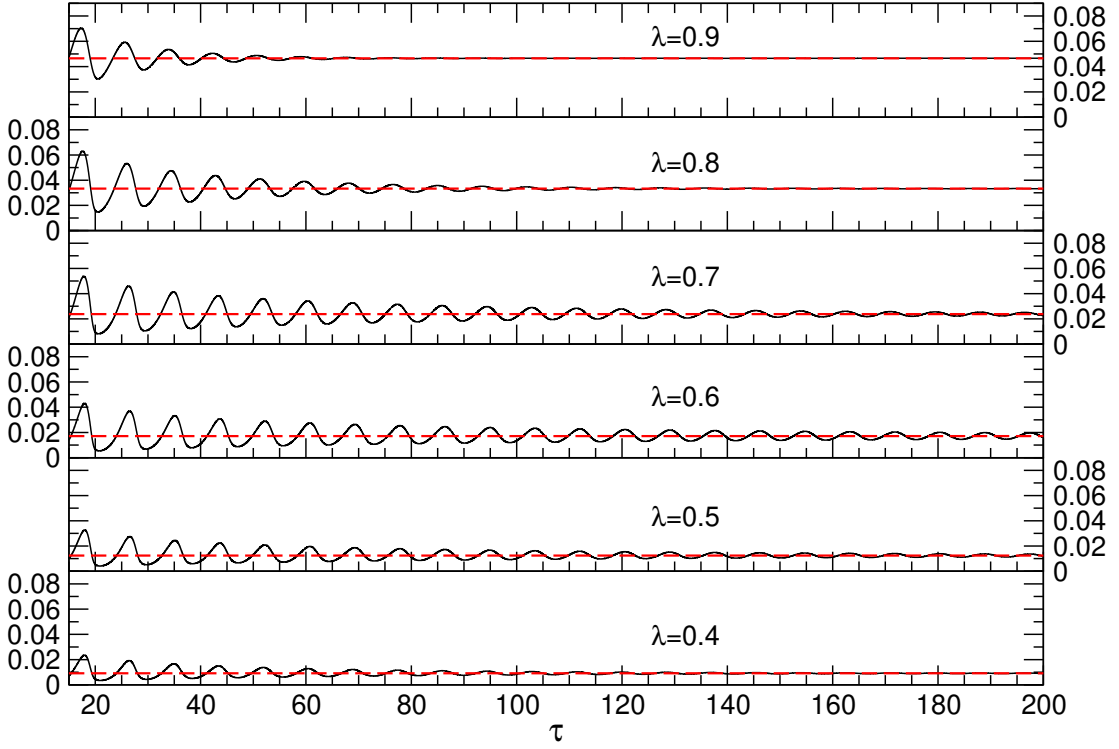}
\caption{\label{fig_oscN10000}
Time evolution of $\vk_1(\tau)$ for $\mu=10^{-5}$, $N=10000$, and several values of $\lambda$. The oscillations observed in figure \ref{fig_oscN1000} disappear, and the {\tc constant}
value $\vk_1^*$ (dashed lines) is reached asymptotically.}
\end{figure*}

Plotted in figures \ref{fig_oscN1000} and \ref{fig_oscN10000} are the time-dependent solutions of \erefs{eq_vk_1}, obtained via the Runge-Kutta-Fehlberg (RKF45) algorithm, for $\mu=10^{-5}$ and two different values of $N$. For small $N\,(=1000)$ steady oscillations are observed in a window around $\lambda\simeq 0.6$
whereas for larger $N\,(=10000)$ these oscillations are transient and tend to disappear, as $\vk_1(\tau)$ approaches its limiting value given by \eref{eq_v1eq2}. This means that for finite $N$ the system governed by \erefs{eq_vk_1},
which conserve the total mass, displays steady oscillations depending on $N$. In the limit $N\rightarrow\infty$, however, {\tc the system does not sustain these oscillations and displays
only time-independent solutions}.

\begin{figure*}[ht]
\centering
\subfloat[]{
\includegraphics[angle=0,scale=0.4,clip]{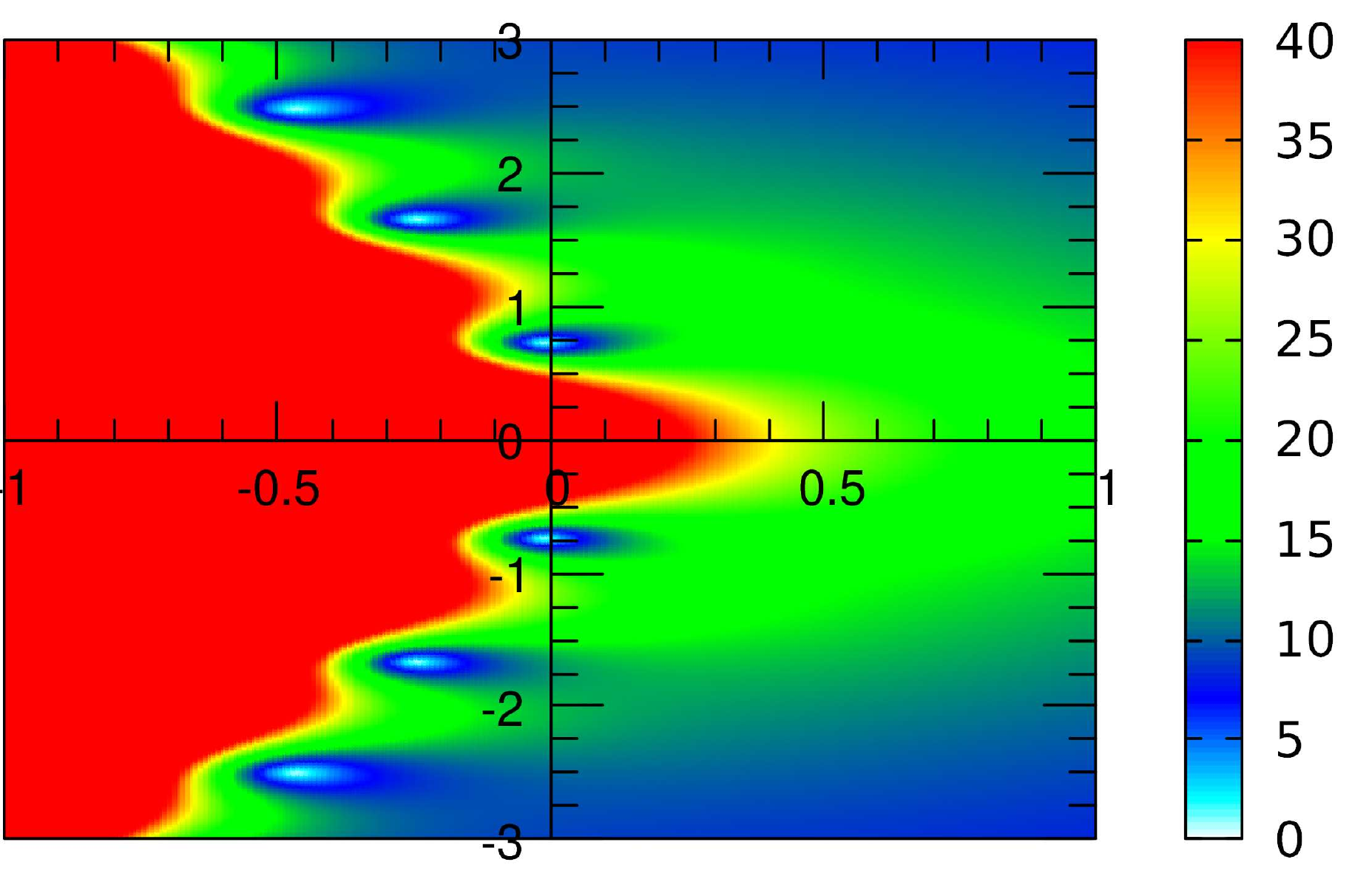}
}
\subfloat[]{
\includegraphics[angle=0,scale=0.4,clip]{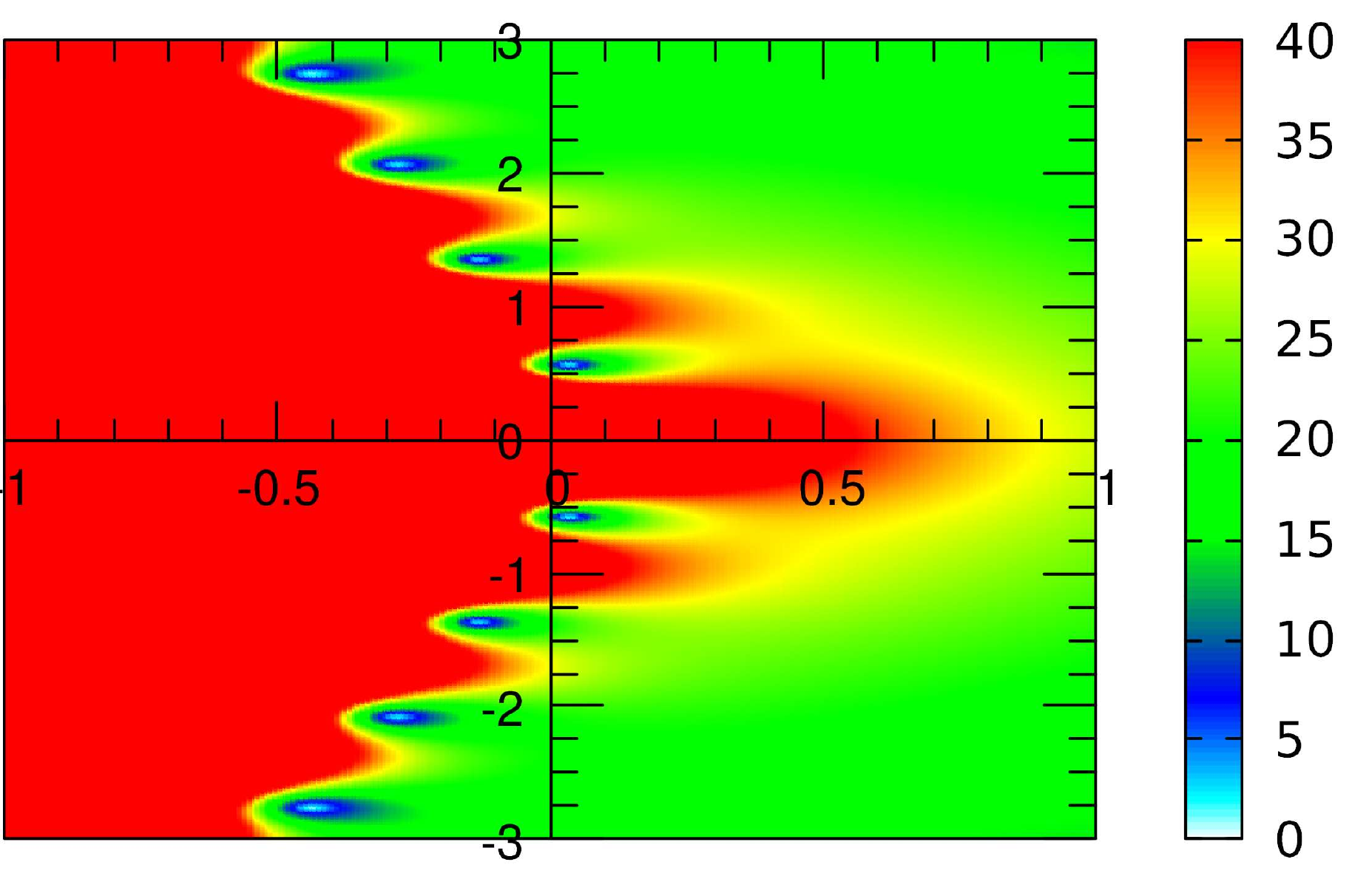}
}
\caption{\label{fig_surf1}
Surface plots of the modulus of \eref{eq_Omega_J1} in the complex $\Omega$-plane $(\Re\,\Omega,\Im\,\Omega)$, {\tc for $\lambda=0.6$ and $\mu=10^{-5}$ (a) and $\mu=10^{-6}$ (b)}. In (a), all the zeros are located on the left part of the plane with
negative real parts and therefore {\tc $\vk_1(\tau)$ tends asymptotically to the time-independent solution $\vk_1^*$. Transient oscillations can occur in the case where the zeros are close to the vertical axis with a typical lifetime inversely proportional to the damping factor}. In (b), one conjugate pair of zeros has a small positive real part, giving rise to {\tc self-generated} oscillations.}
\end{figure*}

\begin{figure*}[ht]
\centering
\subfloat[]{
\includegraphics[angle=0,scale=0.3,clip]{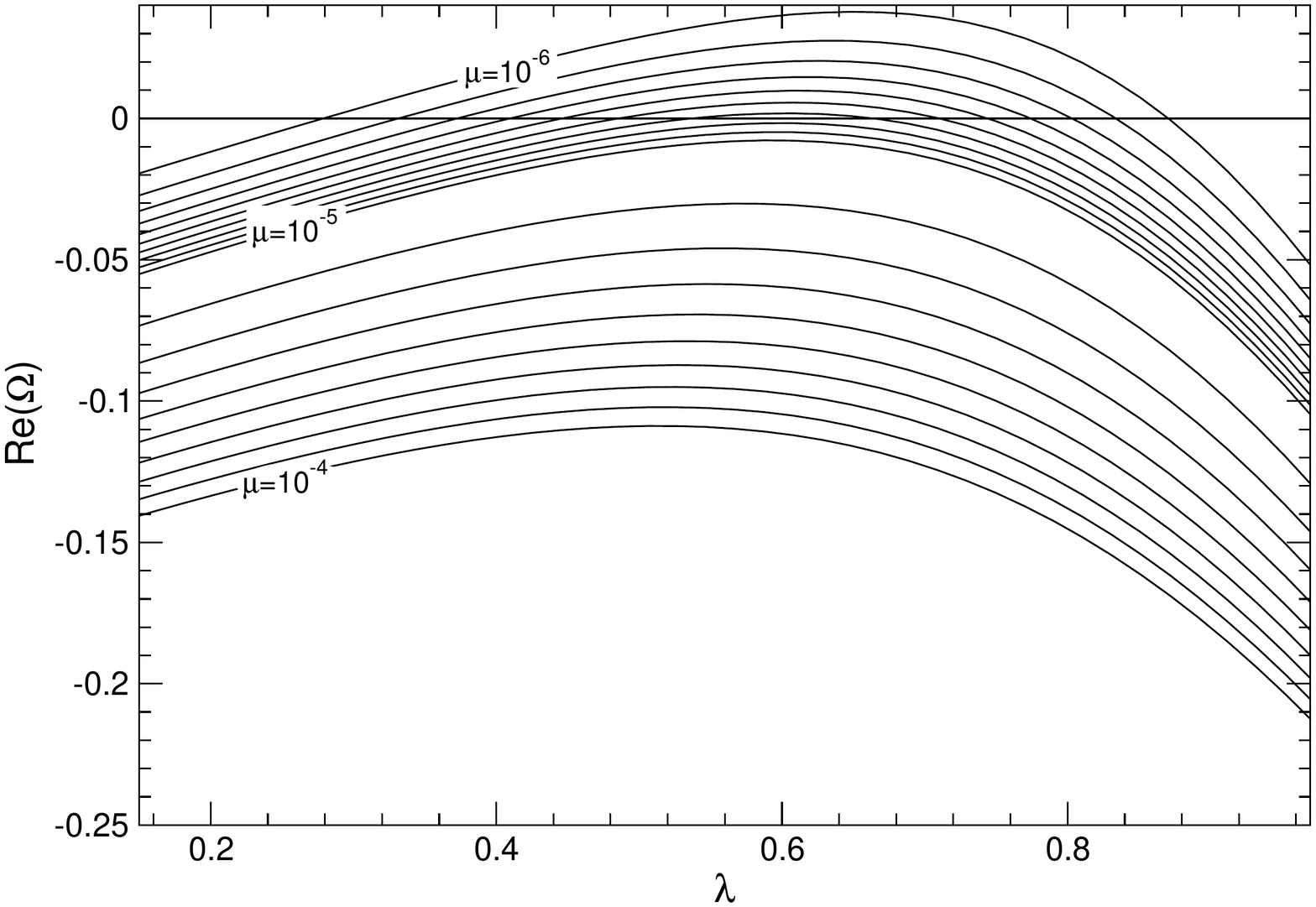}
}
\subfloat[]{
\includegraphics[angle=0,scale=0.3,clip]{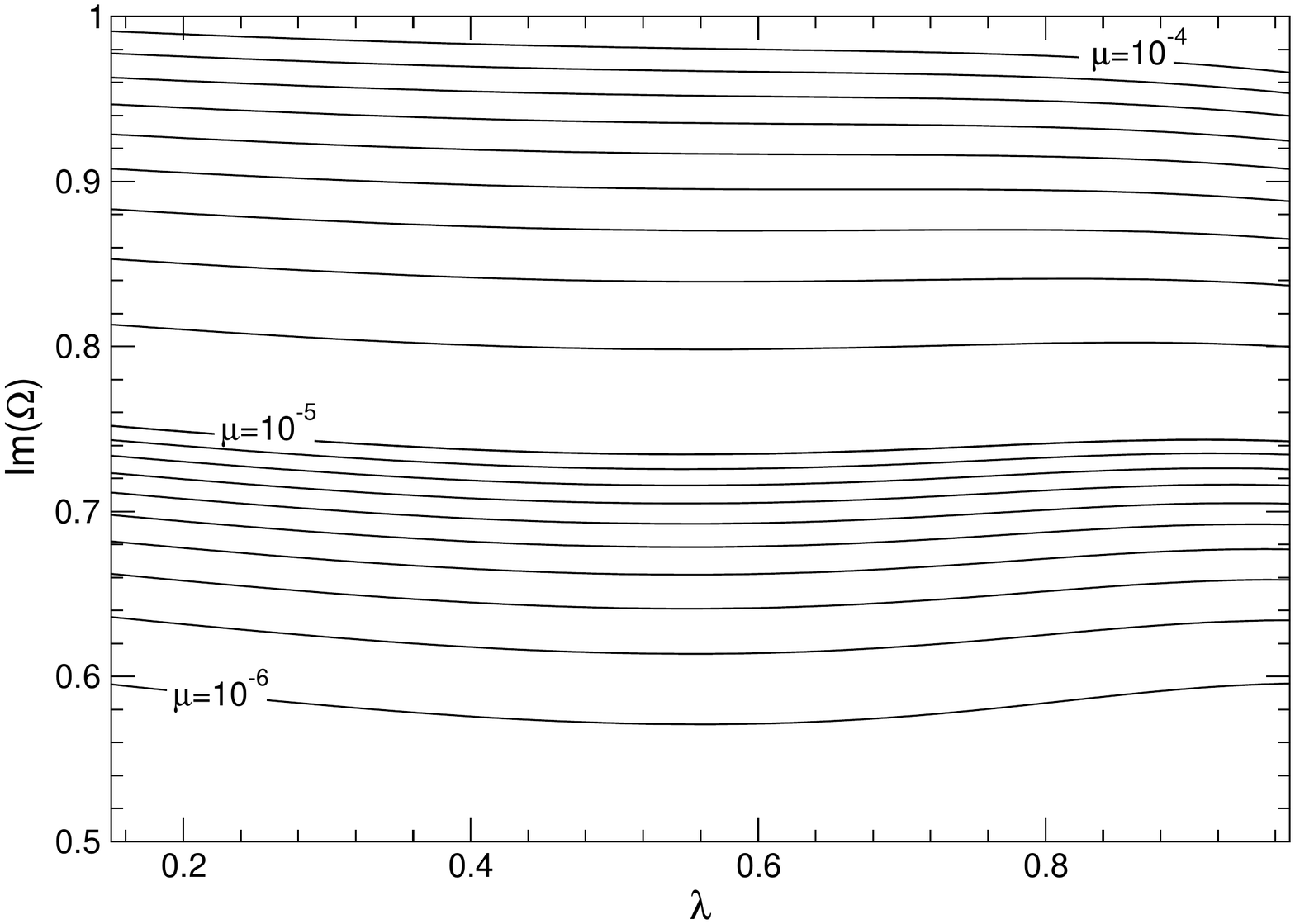}
}
\caption{\label{fig_omega1}
(a) Plots of (a) the real part and (b) the imaginary part of the solution of \eref{eq_Omega_J1} having the largest real part versus $\lambda$, for various values of $\mu$.
}
\end{figure*}

In order to examine the {\tc existence and stability of the self-generated oscillations, solutions of the nonlinear set of dynamical \erefs{eq_vk_1}}, we consider a perturbation $\epsilon(\tau)$ {\tc of the density $\vk_1(\tau)$ around the time-independent solution $\vk_1^*$, such that} $\vk_1(\tau)=\vk_1^*+\epsilon(\tau)$. Then linearization of \eref{eq_v1} yields the integral equation for $\epsilon(\tau)$:
\bb \fl \nn
\epsilon(\tau)&=& -2
\int_0^{\tau} d\tau' \frac{\epsilon(\tau')e^{-(2+\lambda+\mu/\vk_1^*)(\tau-\tau')}}{\left(
1-\int_{0}^{\tau-\tau'}d\tau'' e^{-(1+\mu/\vk_1^*)\tau''}
\right)^3}
-\frac{2\mu}{\vk_1^*}\int_0^{\tau}d\tau'
\frac{e^{-(2+\lambda+\mu/\vk_1^*)(\tau-\tau')}}{\left(
1-\int_{0}^{\tau-\tau'}e^{-(1+\mu/\vk_1^*)\tau''}d\tau''
\right)^3}
\\ \label{eq_eps1}
&& \times \left[
2\int_{\tau'}^{\tau}d\tau'' \epsilon(\tau'')
+
\frac{3\int_{\tau'}^{\tau}d\tau'' e^{-(1+\mu/\vk_1^*)(\tau-\tau'')}\int_{\tau''}^{\tau}d\tau'''
\epsilon(\tau''')}{\left (1-\int_{0}^{\tau-\tau'}d\tau'' e^{-(1+\mu/\vk_1^*)\tau''}\right )}
\right].
\ee
Note that the {\tc above formula} does not take the form of an expansion in $\mu$.
Since there is no obvious convolution integral {\tc in this expression}, we can not take a Laplace transform but consider a perturbation in the exponential form: $\epsilon(\tau)=\epsilon_0e^{\Omega\tau}$ with $\epsilon_0$ being a small
amplitude {\tc and $\Omega$ a complex number whose imaginary part corresponds to a frequency and real part to a damping factor which can be positive, in this case the perturbation is relevant, or negative  and the perturbation is irrelevant. We want therefore to analyze the existence of growing perturbations
leading to self-oscillations if the imaginary part of $\Omega$ is non-zero and real part positive.} After some algebra, we obtain the equation for the complex frequency $\Omega$ {\tc which is exact in the asymptotic limit $N\rightarrow\infty$:}
\bb\nn
1&+&2\fJ_3(\beta{+}\Omega,\alpha)+\frac{2\mu}{\Omega\vk_1^*}\Big(
2\Big[\fJ_3(\beta,\alpha)-\fJ_3(\beta{+}\Omega,\alpha)\Big]
\\ \label{eq_Omega_J1}
&+&\frac{3}{\alpha}\Big[\fJ_4(\beta,\alpha)-\fJ_4(\beta{+}\alpha,\alpha) \Big]
-\frac{3}{\alpha+\Omega}\Big[\fJ_4(\beta,\alpha)-\fJ_4(\beta{+}\alpha{+}\Omega,\alpha)\Big]
\Big)=0 ,
\ee
where $\alpha \equiv 1+\mu/\vk_1^*$, $\beta \equiv 2\alpha+\lambda$, and
%
\bb \fl \nn
\fJ_k(\beta,\alpha) &\equiv& \alpha^k\int_0^{\tau}d\tau' \frac{e^{-\beta\tau'}}{\left (\alpha-1+e^{-\alpha\tau'}\right )^k}
\simeq \frac{\alpha^{k-1}}{(\alpha-1)^{k-\beta/\alpha}}\int_0^{(\alpha-1)^{-1}} dx
\frac{x^{\beta/\alpha-1}}{(1+x)^k}
\\ \nn
&=& \left (\frac{\alpha}{\alpha-1}\right)^{k-1}
\left [(-1)^{k+1}
\frac{\pi\Gamma(\beta/\alpha)(\alpha-1)^{\beta/\alpha-1}}{\Gamma(k)\Gamma(1{+}\beta/\alpha{-}k)
\sin(\pi\beta/\alpha)}
\right .
\\ \label{eq_Jk} & &~~~~~~~ \left .
-\frac{\alpha(\alpha-1)^{k-1}{}_2F_1(k,k{-}\beta/\alpha; 1{-}\beta/\alpha{+}k; 1{-}\alpha)}{k\alpha-\beta}
\right ].
\ee
Here the second integral has been obtained in the limit $\tau\rightarrow\infty$. Accordingly, this approximation should be valid {\tc if the following condition holds:} $|\Omega|\tau\gg 1$. The function $\fJ_k$ satisfies the following recursion relation
\bb\nn
\fJ_{k+1}(\beta,\alpha)=-\frac{1}{k}+\frac{\beta-\alpha}{k}\fJ_{k}(\beta{-}\alpha,\alpha) ~~~(\beta>\alpha>1),
\\
\fJ_k(\alpha,\alpha)=\frac{1}{k-1}\left[\left (\frac{\alpha}{\alpha-1}\right )^{k-1}-1 \right] ~~~(k>1,\;\alpha>1) .
\ee
{\tc For illustration,} figure \ref{fig_surf1} presents the modulus of \eref{eq_Omega_J1} for $\lambda=0.6$ {\tc and $\mu=10^{-5}$ (a), and $\mu=10^{-6}$ (b)}. The zeros are distributed on the left part of the complex plane, with negative real parts, except for the case (b), where the rightmost pair of complex conjugate has a slightly positive real part.
Plotted in figure \ref{fig_omega1} is the solution that has the largest
real part for various values of $\mu$, as a function of $\lambda$.
For $\mu$ smaller than $8\times 10^{-6}$, the real part becomes positive around
$\lambda\simeq 0.6$, and oscillations occur with the frequency given by the imaginary
part of the solution, see figure \ref{fig_omega1}(b).

\begin{figure*}[ht]
\centering
\includegraphics[angle=0,scale=0.6,clip]{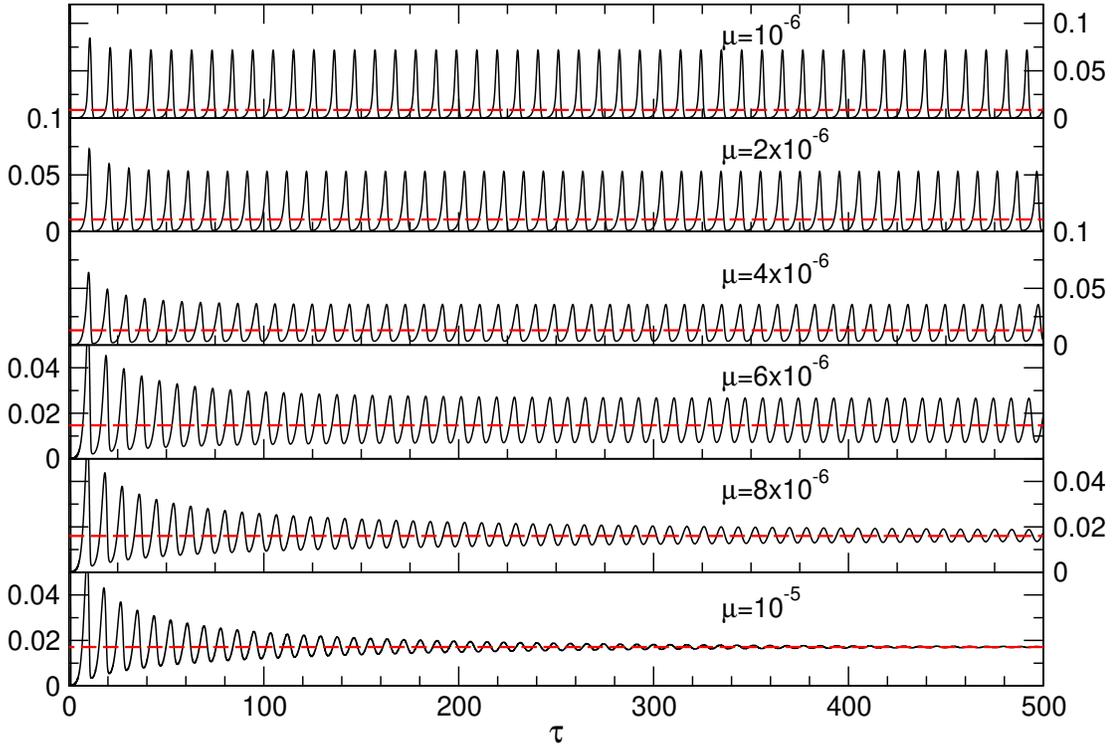}
\caption{\label{fig_osc_mu}
{\tc Numerical solutions of \erefs{eq_vk_1}} for $\lambda=0.6$ and several values of $\mu$ {\tc with $N=10000$}. For $\mu=10^{-5}$ and $8\times 10^{-6}$, oscillations are damped while $\vk_1$ approaches the limit given by \eref{eq_v1eq2}
(red dashed lines).
}
\end{figure*}
%
\begin{figure*}[ht]
\centering
\includegraphics[angle=0,scale=0.6,clip]{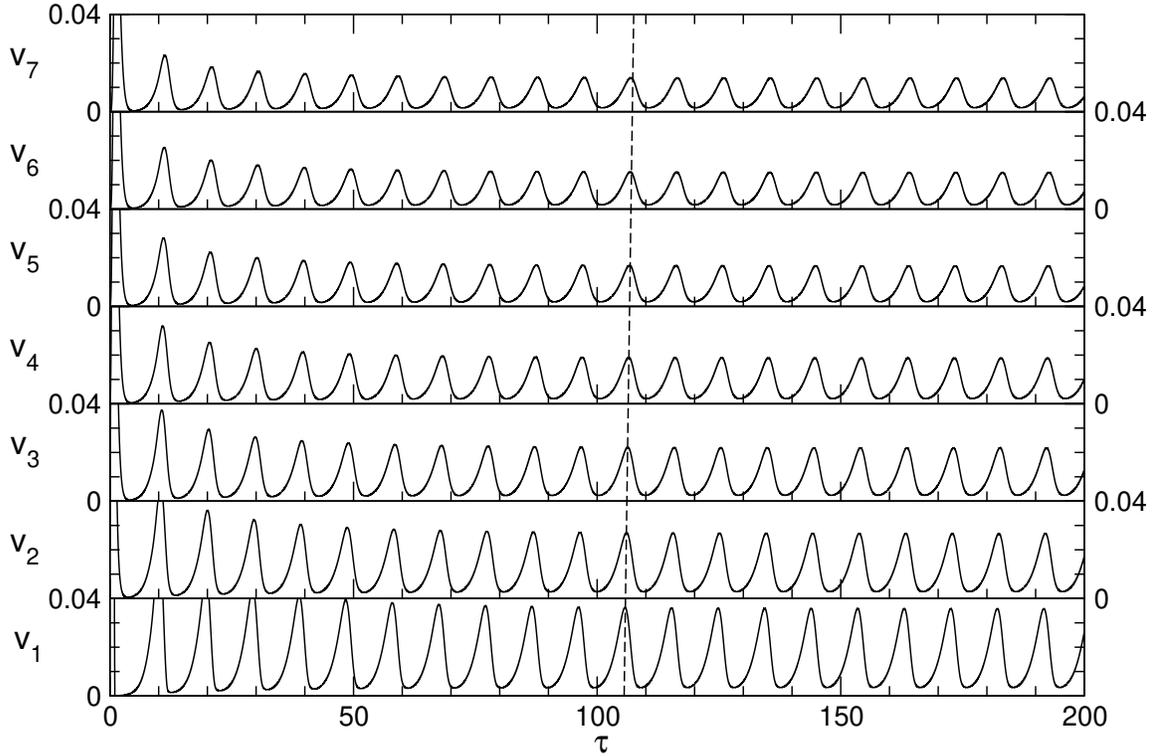}
\caption{\label{fig_vk}
{\tc Numerical solutions of \erefs{eq_vk_1} for $\lambda=0.6$, $\mu=4\times 10^{-6}$,
and with $N=10000$ clusters. We have represented the first reduced densities $\vk_k$, $1\le k\le 7$,
as an illustration. The amplitude of the oscillations decreases with the cluster size (the scale of the vertical axis is the same for all the $\vk_k$s).
The dashed line is a guide to visualize the shift between corresponding peaks.} 
}
\end{figure*}
%

\begin{figure*}[ht]
\centering
\subfloat[]{
\includegraphics[angle=0,scale=0.4,clip]{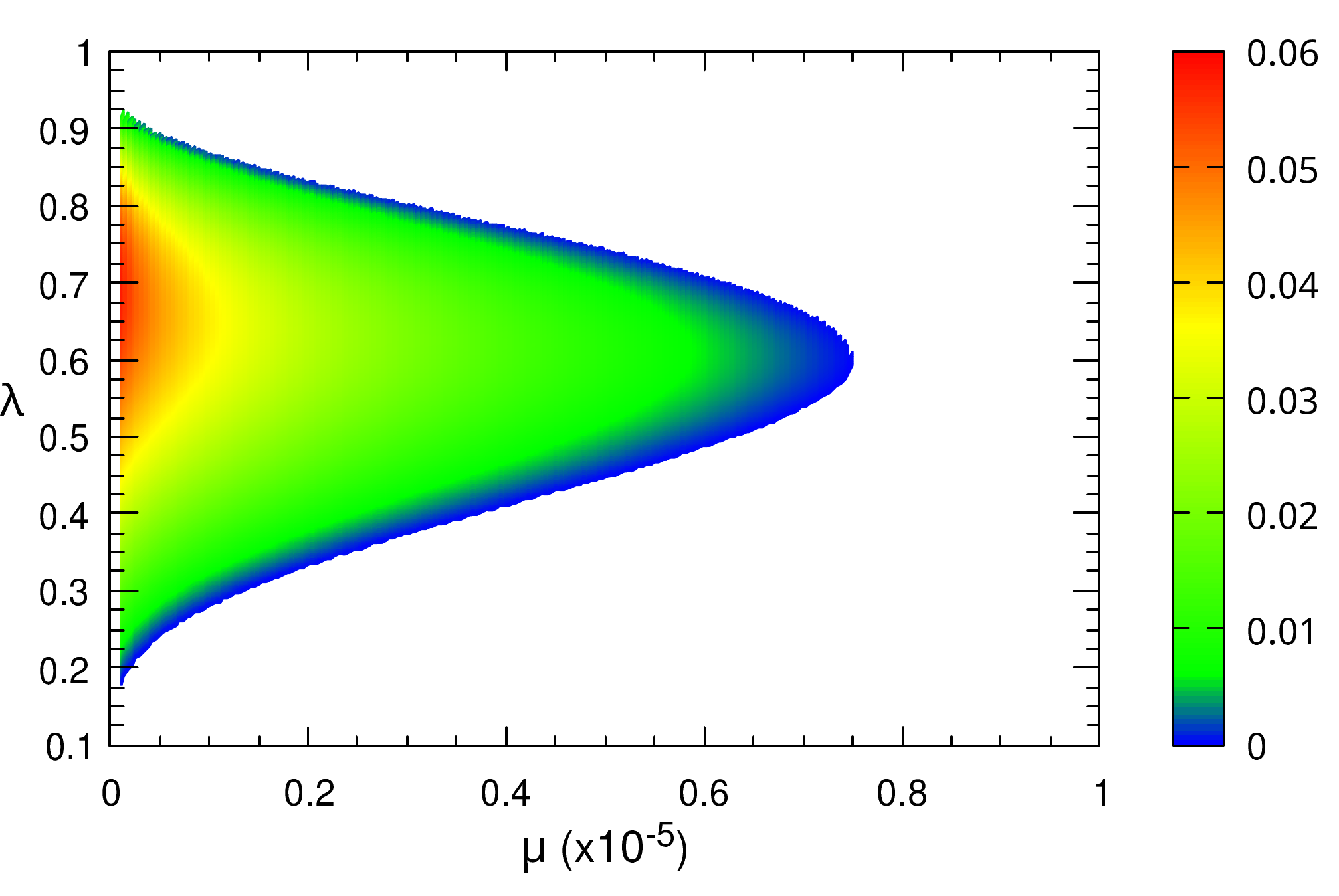}
}
\subfloat[]{
\includegraphics[angle=0,scale=0.4,clip]{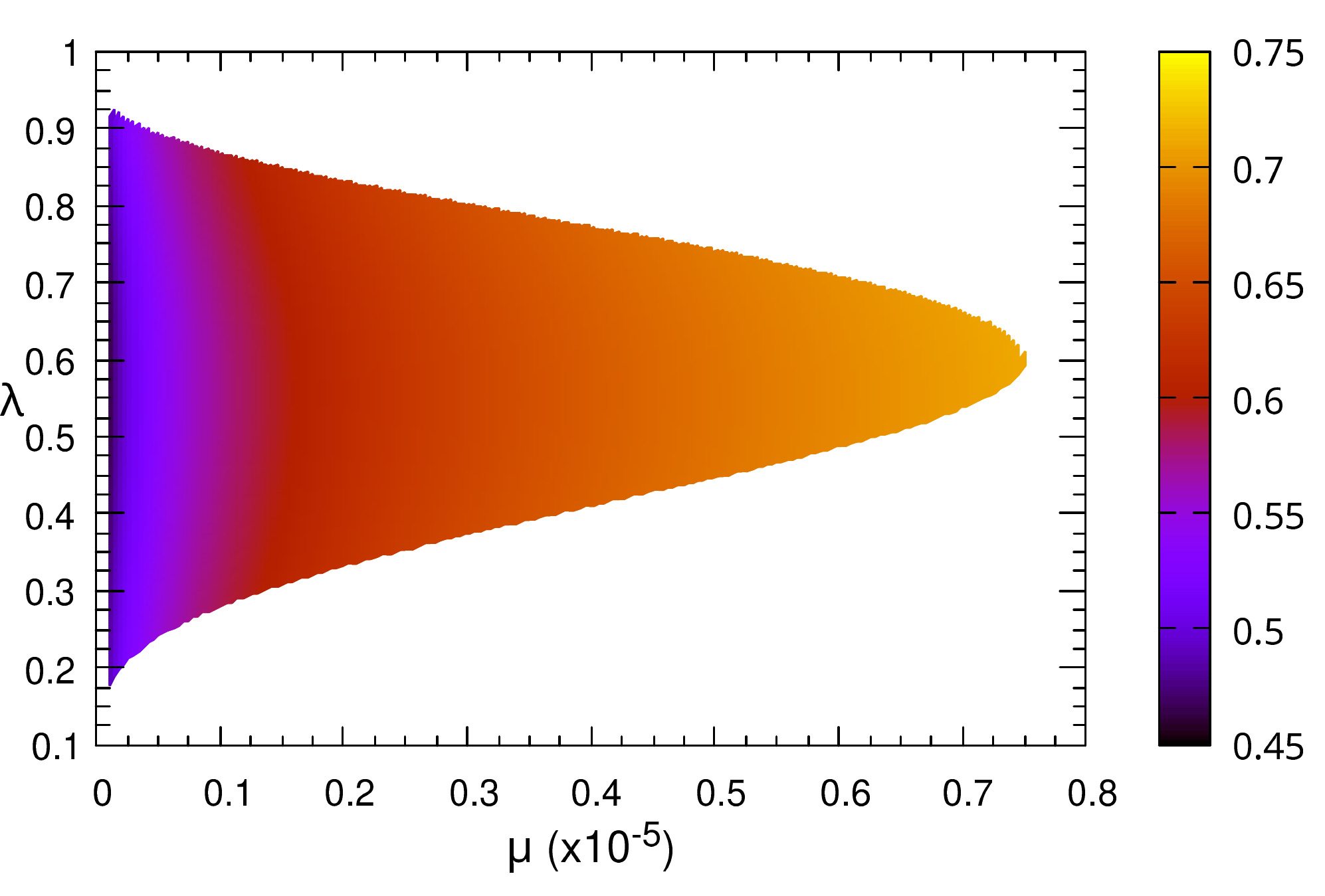}
}
\caption{\label{fig_omega1_surf}
(a) Plots of the real (a) and imaginary (b) parts of the solution of \eref{eq_Omega_J1} with the largest real part, providing the phase diagram in the plane {\tc $(\mu,\lambda)$}.
}
\end{figure*}
%
{\tc In figure \ref{fig_osc_mu}, we have numerically solved \erefs{eq_vk_1} for $N=10000$ using the RKF45 algorithm, and steady oscillations can be seen for $\lambda=0.6$ and for small values of $\mu$ in the range $0<\mu<8\times 10^{-6}$}. For $\mu=8\times 10^{-6}$, oscillations are slowly damped as $\tau$ increases, since the real part of the corresponding $\Omega$ is negative but close to zero. {\tc In figure \ref{fig_vk} is illustrated the oscillation behavior of $\vk_k$ for the first seven cluster sizes $k$. We have
taken $(\mu,\lambda)=(4\times 10^{-6},0.6)$. The amplitude of $\vk_k$ decreases with the cluster size.}

Figure \ref{fig_omega1_surf} the phase diagram in the plane {\tc $(\mu,\lambda)$}
corresponding {\tc to the existence of steady oscillations (domain in color). It is obtained by
evaluating whether there is at least one pair of conjugate solutions $\Omega$ from figure \ref{fig_surf1} with a strictly positive real part, see figure \ref{fig_omega1_surf}(a). Figure \ref{fig_omega1_surf}(b) shows the corresponding positive imaginary part or oscillation frequency.} The domain is bounded by $0<\lambda<1$ and $0<\mu<8\times 10^{-6}$.
{\tc The case $\mu=0$ is excluded even if the range $0<\lambda<1$ is part of the oscillatory domain 
where at least one solution $\Omega$ presents a positive real part. Indeed, $\vk_1(\tau)$ oscillates but takes
negative values and therefore the dynamics stops at a finite time as no one particle cluster
is left in the system. The role of $\mu\neq 0$ is 
to prevent $\vk_1$ to become negative by strong nonlinear effects when $\vk_1$ approaches zero, no matter how small its value is \cite{Fortin:2022}.}
Carrying out the Fourier transform of the signals in figure \ref{fig_osc_mu} for values of $\mu$ {\tc in the domain of oscillations}, we can extract the frequency $\Omega_{\textrm{num}}$ and present the obtained values in Table \ref{tab:table1}.
These values are compared with the imaginary parts of the solutions of \eref{eq_Omega_J1}. It is observed that the differences are small when $\mu$ is close to the oscillation threshold. They tend to increase when $\mu$ is decreased or the oscillation amplitude is increased. 

%
\begin{table}[h]
\centering
\caption{Frequency $\Omega_{\textrm{num}}$ for $\lambda=0.6$ and several values of $\mu$, extracted from the Fourier transform of the time oscillations in figure \ref{fig_osc_mu}. For comparison, the imaginary part $\Im \Omega$ of the solution of \eref{eq_Omega_J1} is also given.}
\label{tab:table1}
{\renewcommand{\arraystretch}{1.2}
\begin{tabular}{c|c|c}
$\mu$ & $\Omega_{\textrm{num}}$ & $\Im\,\Omega$
\\  \hhline{===}
$1\times 10^{-6}$ & 0.571 &  0.601
\\
$2\times 10^{-6}$ & 0.614 & 0.620
\\
$4\times 10^{-6}$ & 0.662 & 0.658
\\
$6\times 10^{-6}$ & 0.692 & 0.690
\\
\hline
\end{tabular}}
\end{table}
%

\subsection{Model with $(\as,\bs)=(1,1)$ \label{sec_11}}
%
In this section, we consider the case $(\as,\bs)=(1,1)$. As before, \erefs{eq_gen_vk} lead
to a linear PDE for the generating function $G(u,\tau)$, similar to \eref{eq_G}:
\bb\label{eq_G2}
\frac{\partial G(u,\tau)}{\partial\tau}+(1+\lambda+\mu\vk_1^{-1}-e^u)\frac{\partial G(u,\tau)}{\partial u}
=e^uG(u,\tau)+2e^{2u}\vk_1 .
\ee
The solution can be obtained via parametrizing $u$ and $\tau$ such that $\tilde G(s)\equiv G(u(s),\tau(s))$
with
\bb
\tau'(s)=1 ~~\mbox{and} ~~ u'(s)=1+\lambda+\frac{\mu}{\vk_1(\tau(s))}-e^{u(s)},
\ee
which is integrated into
\bb \fl 
\tau(s)&=&s, \\ \fl\nn
u(s)&=&\int_0^s ds' \left(1+\lambda+\frac{\mu}{\vk_1(s')}\right)
-\log\left( C+\int_0^s ds' \exp\Big [\int_0^{s'}ds'' (1+\lambda+\mu\vk_1^{-1}(s''))\Big ]\right).
\ee
Here $C$ is a constant determined by the condition that the
point $(u,\tau)$ belongs to the curve, namely, $s=\tau$ and $u(\tau)=u$, which gives
\bb
C=\exp\left(\gamma(\tau)-u\right) - \int_0^{\tau} ds \exp\left(\gamma(s)\right),
\ee
with $\gamma(s) = (1+\lambda)s+\int_0^{s} ds' \mu\vk_1^{-1}(s')$.
The PDE (\ref{eq_G2}) therefore reduces to
\bb \label{eq_Gs2}
\tilde G'(s)
=e^{u(s)}\tilde G(s)+2e^{2u(s)}\vk_1(s) ,
\ee
which can be integrated into
\bb\nn
\tilde G(s)=2e^{W(s)}\int_0^s ds' \vk_1(s')e^{-W(s')+2u(s')},
\\
W(s)=\int_0^s ds' e^{u(s')}= \log\left(C+\int_0^s ds' e^{\gamma(s')}\right)-\log C .
\ee
Simplifying and using $G(0,\tau)=1-\vk_1(\tau)$, we obtain
an implicit integral equation for $\vk_1(\tau)$:
\bb \label{eq_v1_2}\fl
\vk_1(\tau)=1-2
\int_0^{\tau} d\tau'
\frac{\vk_1(\tau')\exp\left[-2(1+\lambda)(\tau-\tau')-2\mu\int_{\tau'}^{\tau}d\tau'' \vk_1^{-1}(\tau'')\right]}
{\left (1-\int_{\tau'}^{\tau} d\tau'' \exp\left[-(1+\lambda)(\tau-\tau'')-\mu\int_{\tau''}^{\tau} d\tau''' \vk_1^{-1}(\tau''')\right]
\right )^3} ,
\ee
which is similar to \eref{eq_v1} except for the extra $\lambda$ term in the exponential
rates.
\subsubsection{{\tc Solution in the case $\mu=0$.}\label{subsubsect_mu0}}
In the case $\mu=0$, \eref{eq_v1_2} {\tc yields the more simple integral equation}
\bb \label{eq_v1_3}\fl
\vk_1(\tau)=1-2
\int_0^{\tau} d\tau' \,\frac{\vk_1(\tau')e^{
-2(1+\lambda)(\tau-\tau')}}{\left(
1-\int_{0}^{\tau-\tau'}e^{-(1+\lambda)\tau''}
d\tau''\right)^3},
\ee
which can be solved by means of a Laplace transform:
\bb
\hat\vk_1(p)=\frac{1}{p}\frac{1}{1+2\hat F(p)},
\mathrm{with}\, 
\hat F(p)=(1+\lambda)^2\lambda^{p/(1+\lambda)-1}\int_0^{1/\lambda}dx
\frac{x^{1+p/(1+\lambda)}}{(1+x)^{3}}.
\ee
This results in the expression
\bb \label{eq_v1_mu0_L}
\hat\vk_1(p)=
\frac{\sin\left(\pi\frac{p}{\lambda+1}\right)}{p(p+\lambda+1)\left(
\pi p\lambda^{\frac{p}{\lambda+1}-1}
-\sin\left(\pi\frac{p}{\lambda+1}\right)\left[p\Phi\left(-\lambda,1,1{-}\frac{p}{\lambda+1}\right)+1\right]\right)},
\ee
where the Lerch transcendent function $\Phi$ is defined by the formal series $\Phi(z,s,a)=\sum_{k\ge 0} z^k (k+a)^{-s}$. This solution possesses a long-time limit $\hat\vk_1(p)\simeq p^{-1}\lambda/(1+\lambda)$ as
$p\rightarrow 0$, and therefore $\vk_1(\tau)$ approaches the {\tc time independent value} $\vk_1^*=\lambda/(1+\lambda)$. However, when $\lambda$ is {\tc smaller than a threshold value $\lambda_c$}, {\tc we observe that} $\vk_1(\tau)$ becomes negative before reaching the asymptotic value $\vk_1^*$. By solving numerically the system \eref{eq_gen_vk} for, e.g., $N=10000$, {\tc we find that 
$\lambda_c\simeq 0.16775$ and the dynamics stops at a finite time $\tau$ when $\vk_1$ vanishes before becoming negative.} On the other hand, for $\lambda>\lambda_c$, the densities reach the time-independent equilibrium state where $\vk_1(\tau)=\vk_1^*$, and the other values decrease exponentially as
\bb
\vk_k=\frac{\lambda}{(\lambda+1)^k}\;(k\ge 2),
\ee
see \eref{eq_v1_mu0_Lbis} below.
Another method is to solve \eref{eq_gen_vk} for $(\as,\bs)=(1,1)$ and $\mu=0$. This yields
the following system of {\tc linear} equations:
\bb\nn
&\frac{\parti \vk_1}{\parti\tau}=-2\vk_1-\sum_{k=2}^{\infty}(1-\lambda k)\vk_k,
\\ \nn
&\frac{\parti \vk_k}{\parti \tau}=
k\vk_{k-1}-k(1+\lambda)\vk_k \;(k\ge 2).
\ee
Taking the Laplace transform, we obtain recursively
\bb\nn
\left(\lambda\sum_{k=2}^{\infty}\frac{k!k}{\prod_{j=2}^k [p+j(1+\lambda)]}-1-p\right) \hat\vk_1(p)
=\frac{1}{p}-1,
\\ \label{eq_v1_mu0_Lbis}
\hat\vk_k(p)=\prod_{j=2}^k\left[\frac{j}{p+j(1+\lambda)}\right] \hat\vk_1(p)\;(k\ge 2).
\ee
The summation in the first expression can be reduced to a hypergeometric function as
\bb\nn
\sum_{k=2}^{\infty}\frac{k!k}{\prod_{j=2}^k [p+j(1+\lambda)]} &=&
\sum_{k=1}^{\infty}\frac{\Gamma(k{+}2)\Gamma(k{+}2)}{\Gamma(k{+}2{+}p/(\lambda{+}1))\Gamma(k{+}1)}(\lambda+1)^{-k}
\\
&=&{}_2F_1\left (2,2;2{+}\frac{p}{\lambda{+}1};\frac{1}{\lambda{+}1}\right )-1 ,
\ee
which in turn leads to
\bb
\left[{}_2F_1\left (2,2;2{+}\frac{p}{\lambda{+}1};\frac{1}{\lambda{+}1}\right )\lambda-\lambda-1-p\right] \hat\vk_1(p)=\frac{1}{p}-1 .
\ee
{\tc We have checked numerically that this} is formally equivalent to the expression in \eref{eq_v1_mu0_L} and gives an identity
between the hypergeometric function ${}_2F_1$ and the Lerch function $\Phi$ by eliminating $\hat\vk_1$.
\subsubsection{Stability around the {\tc steady-state} solution {\tc for $\mu=0$}}
\begin{figure*}[ht]
\centering
\subfloat[]{
\includegraphics[angle=0,scale=0.4,clip]{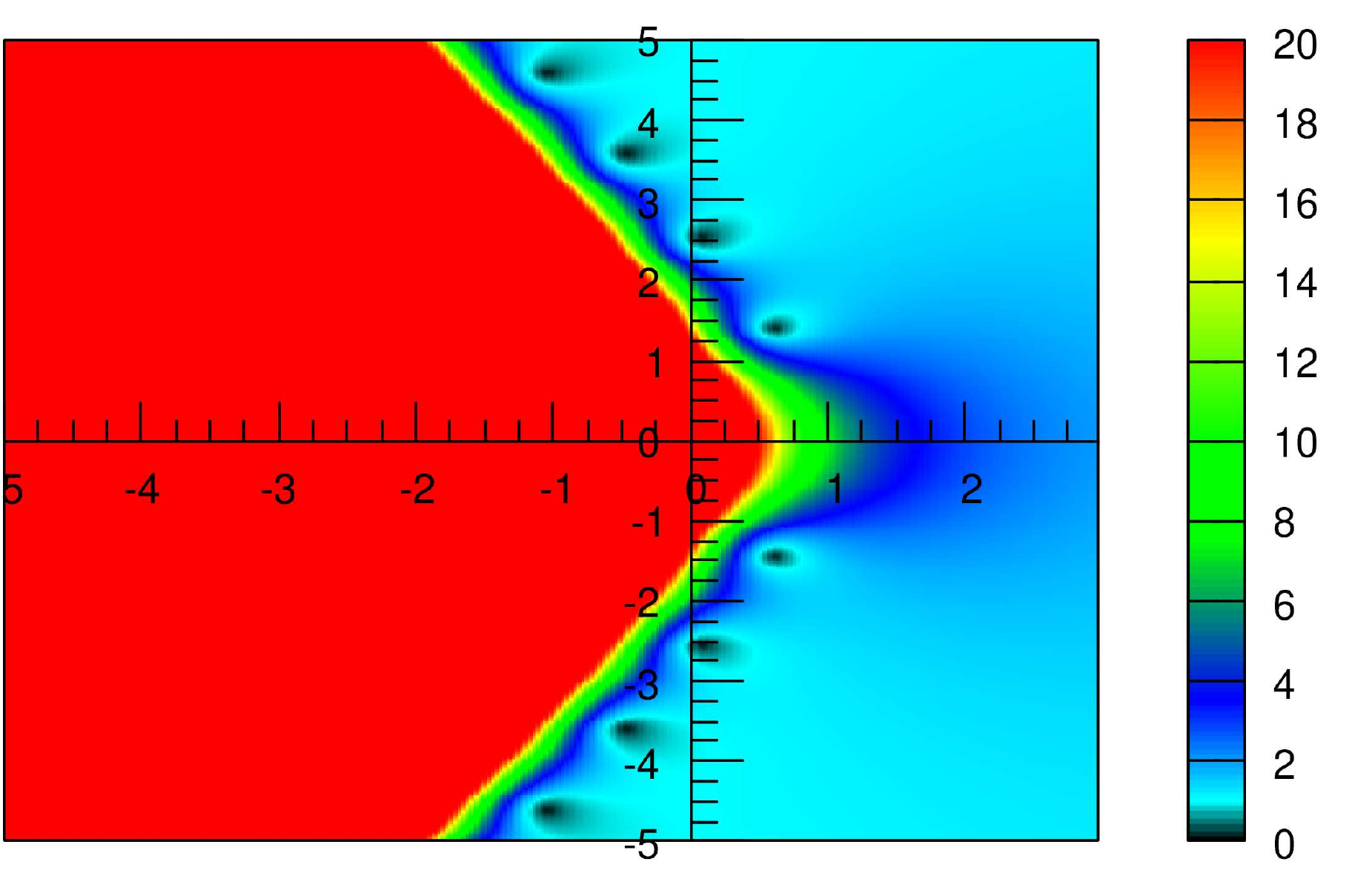}
}
\subfloat[]{
\includegraphics[angle=0,scale=0.3,clip]{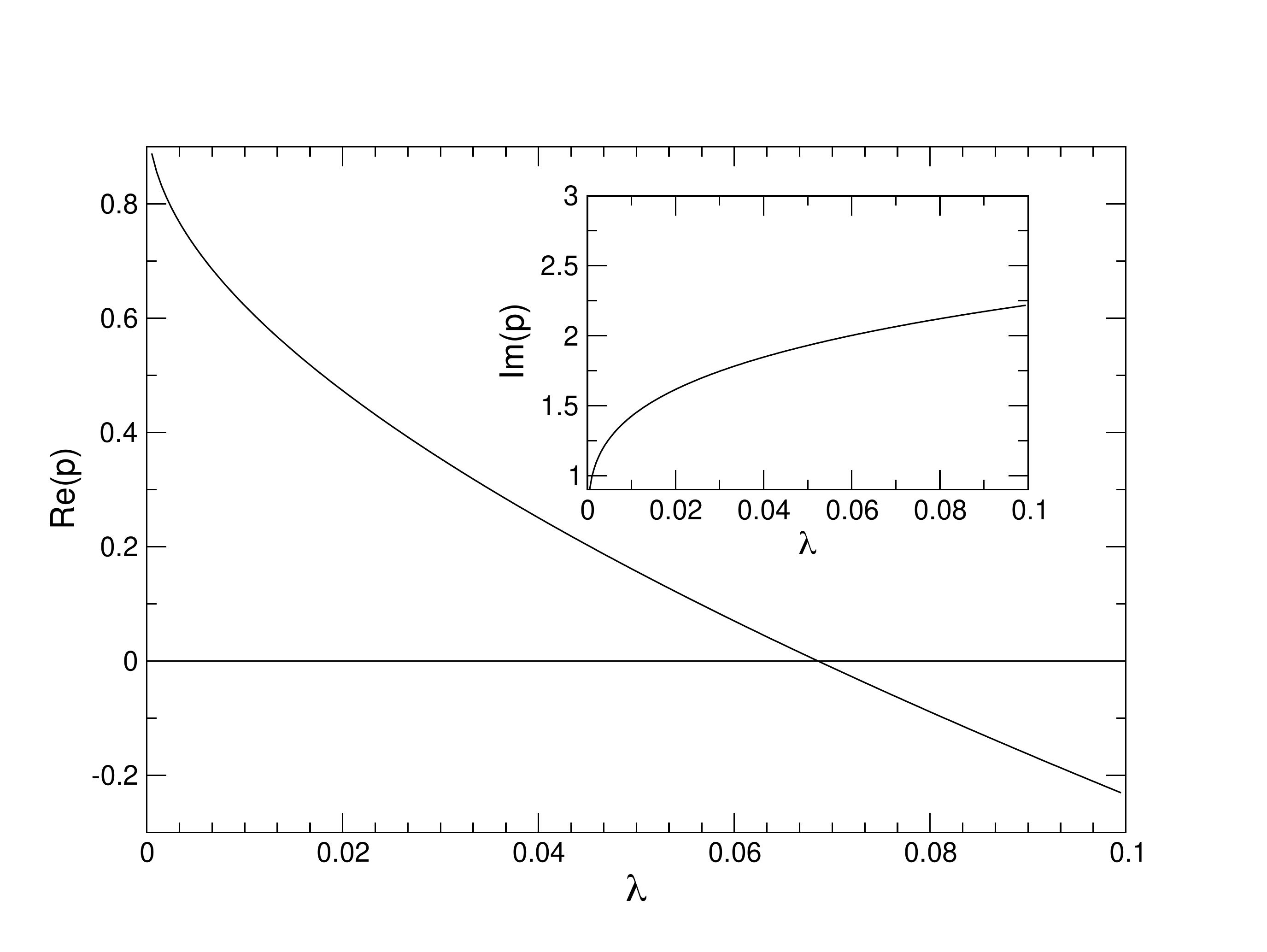}
}
\caption{\label{fig_psol}
(a) Surface plot of the amplitude $|1+2\hat F(p)|$ for $\lambda=0.01$ and $\mu=0$ in the complex plane $(\Re\,p, \Im\,p)$. All amplitudes greater than 20 are colored in red. It is shown that for this
value of $\lambda$, there exist two zeros and their conjugates with positive real parts. (b) Plot of the
complex solution of $1+2\hat F(p)=0$ with the largest real part as a function of $\lambda$.}
\end{figure*}

We consider a {\tc time perturbation} $\epsilon(\tau)$ around the constant solution $\vk_1^*$ of \eref{eq_v1_3}, and set $\alpha\equiv 1+\lambda>1$. We thus write $\vk_1(\tau)=\vk_1^*+\epsilon(\tau)$, where $\epsilon$ satisfies the following integral equation {\tc after linearization}
\bb\nn
\epsilon(\tau)&=& -2
\int_0^{\tau}d\tau' \frac{\epsilon(\tau')e^{-2\alpha(\tau-\tau')}}{\left (
1-\int_{0}^{\tau-\tau'}e^{-\alpha\tau''}d\tau''
\right )^3}
=-2\alpha^3 \int_0^{\tau} d\tau' \frac{\epsilon(\tau')e^{-2\alpha(\tau-\tau')}}{\left(
\alpha-1-e^{-\alpha(\tau-\tau')}
\right)^3}
\\
&=&-2\int_0^{\tau}d\tau' \epsilon(\tau')F(\tau-\tau') .
\ee
The Laplace transform of this expression gives $\hat\epsilon(p)[1+2\hat F(p)]=0$, which indicates that a non-zero
solution for the perturbation is possible if there exists $p$ satisfying 
\bb \label{eq_Fp}
1+2\hat F(p)=\left [\frac{\pi p\lambda^{p/(1+\lambda)-1}}{\sin(\pi p/(1{+}\lambda))}
-1-p\Phi(-\lambda,1,1{-}p/(1{+}\lambda))
\right ](1+p+\lambda) = 0.
\ee
The zeros of the function $1+2\hat F(p)$ can be located by plotting the modulus $|1+2\hat F(p)|$
in the complex plane as in figure \ref{fig_psol}(a). The zeros and their conjugates are aligned
in the negative real part of the plane, except for a few zeros whose real parts are strictly positive.
Plotted in figure \ref{fig_psol}(b) are the real part and the positive imaginary part of the zero which
has the largest real part. The real part is positive for all $\lambda$ less than
$\lambda^*\simeq 0.068538$. In this case the perturbation grows exponentially and oscillates with
a frequency equal to the imaginary part $\Im\,p$. However, since $\lambda^*<\lambda_c$,
the solution $\vk_1$ becomes negative, and the dynamics stops before oscillations occur.
As we will see below, the presence of any small value of $\mu$ will lead to oscillations with a positive value of $\vk_1$.

%
\subsubsection{{\tc Steady-state} solution for $\mu\neq 0$}
Here we consider the long-time constant solution of \eref{eq_v1_2}
by assuming that $\vk_1(\tau)$ approaches $\vk_1^*>0$ in the limit $\tau\rightarrow\infty$. This solution should satisfy
\bb \label{eq_v1_2lim}\fl
\vk_1^*=\left [1+2\int_0^{\infty}d\tau'
\frac{e^{-2(1+\lambda+\mu/\vk_1^{*})\tau'}}{\left(
1-\int_{0}^{\tau'}e^{-(1+\lambda+\mu/\vk_1^{*})\tau''}
d\tau''\right)^3}\right ]^{-1} ,
\ee
which, upon integration, yields the quadratic equation for $\vk_1^*$:
\bb
(1+\lambda){\vk_1^*}^{2}+(\mu-\lambda)\vk_1^*-\mu=0,
\ee
with the unique positive solution
\bb\label{eq_v1star}
\vk_1^*=\frac{\lambda-\mu+\sqrt{(\lambda+\mu)^2+4\mu}}{2(\lambda+1)}.
\ee
{\tc When $\mu=0$, we recover the previous result of section \ref{subsubsect_mu0}.
We should notice that, in the integral (\ref{eq_v1_2}), the presence of any
non-zero $\mu>0$ should prevent the solution $\vk_1(\tau)$ from vanishing since
the equation contains the integral of $\mu/\vk_1(\tau)$, which becomes large if $\vk_1(\tau)\rightarrow 0$ and annihilates the corresponding exponentials. We expect the right hand side of this equation to tend therefore to unity, in contradiction with the left hand side which is equal to $\vk_1\ll 1$. Thus we may deduce from this observation that $\vk_1(\tau)$ can not take too small values and stays positive.}
\subsubsection{Stability around the {\tc steady-state} solution}
{\tc The constant value, \eref{eq_v1star}, is the long time limit solution provided that any perturbation diminishes with time}. As before,
we consider a perturbation $\epsilon(\tau)$ such that $\vk_1(\tau)=\vk_1^*+\epsilon(\tau)$.
After linearizing \eref{eq_v1_2} and set $\alpha\equiv 1+\lambda+\mu/\vk_1^*$, one obtains the following implicit integral equation for $\epsilon(\tau)$:
\bb\nn
\epsilon(\tau)&=& -2
\int_0^{\tau} d\tau' \frac{\epsilon(\tau')e^{-2\alpha(\tau-\tau')}}{\left (
1-\int_{0}^{\tau-\tau'}d\tau'' e^{-\alpha\tau''}
\right )^3}
-2\frac{\mu}{\vk_1^*}\int_0^{\tau} d\tau'
\frac{e^{-2\alpha(\tau-\tau')}}{\left (
1-\int_{0}^{\tau-\tau'}e^{-\alpha\tau''}d\tau''
\right )^3}
\\ \label{eq_eps2}
& & ~~~ \times \left [
2\int_{\tau'}^{\tau} d\tau'' \epsilon(\tau'')
+3
\frac{\int_{\tau'}^{\tau}d\tau'' e^{-\alpha(\tau-\tau'')}\int_{\tau''}^{\tau} d\tau'''
\epsilon(\tau''')}{\left (1-\int_{0}^{\tau-\tau'}d\tau'' e^{-\alpha\tau''}\right )}
\right ].
\ee
As before this integral equation can not be simplified via a Laplace transform. Instead, we consider again a perturbation of the form $\epsilon(\tau)=\epsilon_0 e^{\Omega\tau}$,
where $\epsilon_0$ is a small amplitude and $\Omega$ is the complex frequency.
If the real part is negative, $\Re\,\Omega <0$, the perturbation is irrelevant {\tc and the constant solution is stable}. Otherwise we expect oscillatory behavior with frequency $\Im\,\Omega$.
The equation for the complex parameter $\Omega$ reads
\bb\nn
1+2\fJ_3(2\alpha{+}\Omega,\alpha)+\frac{2\mu}{\Omega\vk_1^*}\Big (
2\Big [\fJ_3(2\alpha,\alpha)-\fJ_3(2\alpha{+}\Omega,\alpha)\Big ]
\\ \label{eq_Omega_J2}
~~+\frac{3}{\alpha}\Big [\fJ_4(2\alpha,\alpha)-\fJ_4(3\alpha,\alpha) \Big ]
-\frac{3}{\alpha+\Omega}\Big [\fJ_4(2\alpha,\alpha)-\fJ_4(3\alpha{+}\Omega,\alpha)\Big ]
\Big )=0 ,
\ee
%
which reduces, via some algebra, to
\bb\nn
\Omega(1+\lambda+\Omega)\Phi(1{-}\alpha,1,1{-}\Omega/\alpha)-
\frac{\pi\Omega(1+\lambda+\Omega)(\alpha-1)^{\Omega/\alpha}}{(\alpha-1)\sin(\pi\Omega/\alpha)}
\\ \label{eq_Omega_zeros2}
~~ +\frac{\Omega^2+(1+\lambda+\alpha)\Omega}{\alpha+\Omega}
+\frac{\alpha(\alpha^2\lambda-(1+\lambda)\alpha+1+\lambda)}{(\alpha-1)^2(\alpha+\Omega)}
=0,
\ee
with $\alpha=1+\lambda+\mu/\vk_1^*$.
When $\mu=0$, or $\alpha=1+\lambda$, \eref{eq_Fp} is recovered. 
%
\begin{figure*}[ht]
\centering
\subfloat[]{
\includegraphics[angle=0,scale=0.4,clip]{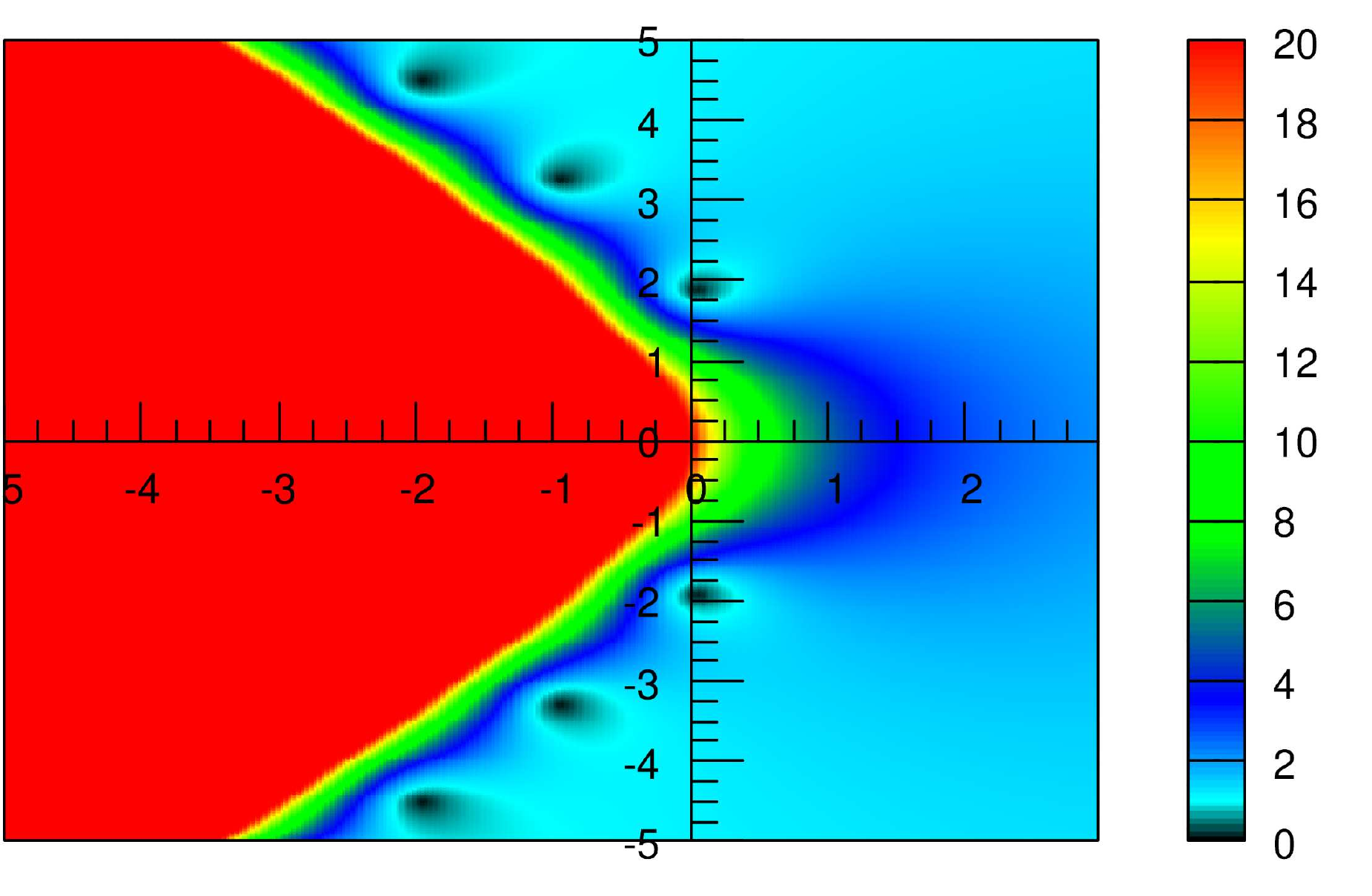}
}
\subfloat[]{
\includegraphics[angle=0,scale=0.4,clip]{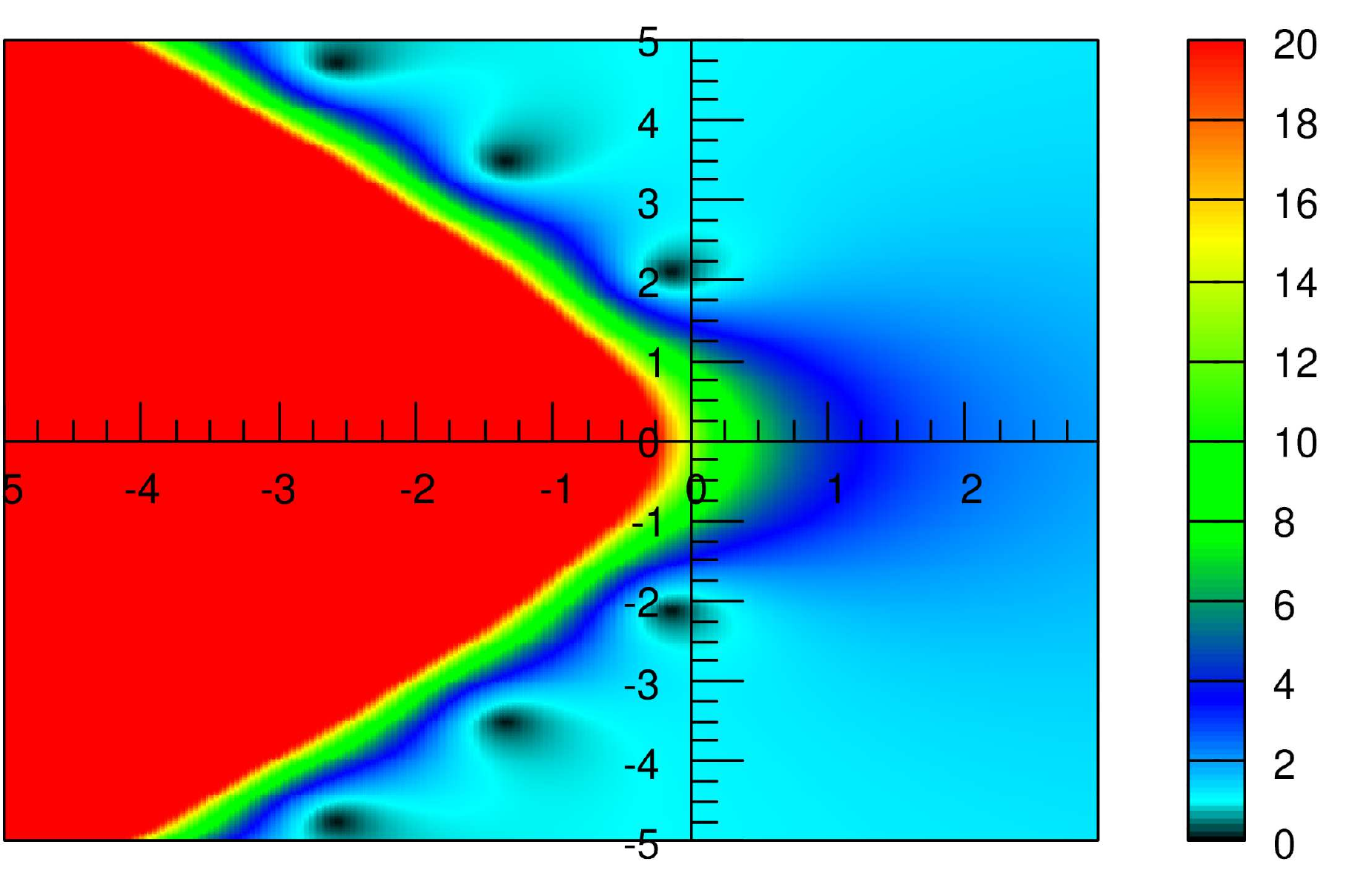}
}
\caption{\label{fig_surf2}
Surface plot of the modulus of \eref{eq_Omega_zeros2} in the complex $\Omega$-plane $(\Re\,\Omega, \Im\,\Omega)$, for $\mu= 10^{-4}$ and $\lambda=0.05$ (a) and $0.08$ (b). In the configuration (a), there exists one conjugate pair of $\Omega$ with a positive real part.}
\end{figure*}
%
\begin{figure*}[ht]
\centering
\subfloat[]{
\includegraphics[angle=0,scale=0.4,clip]{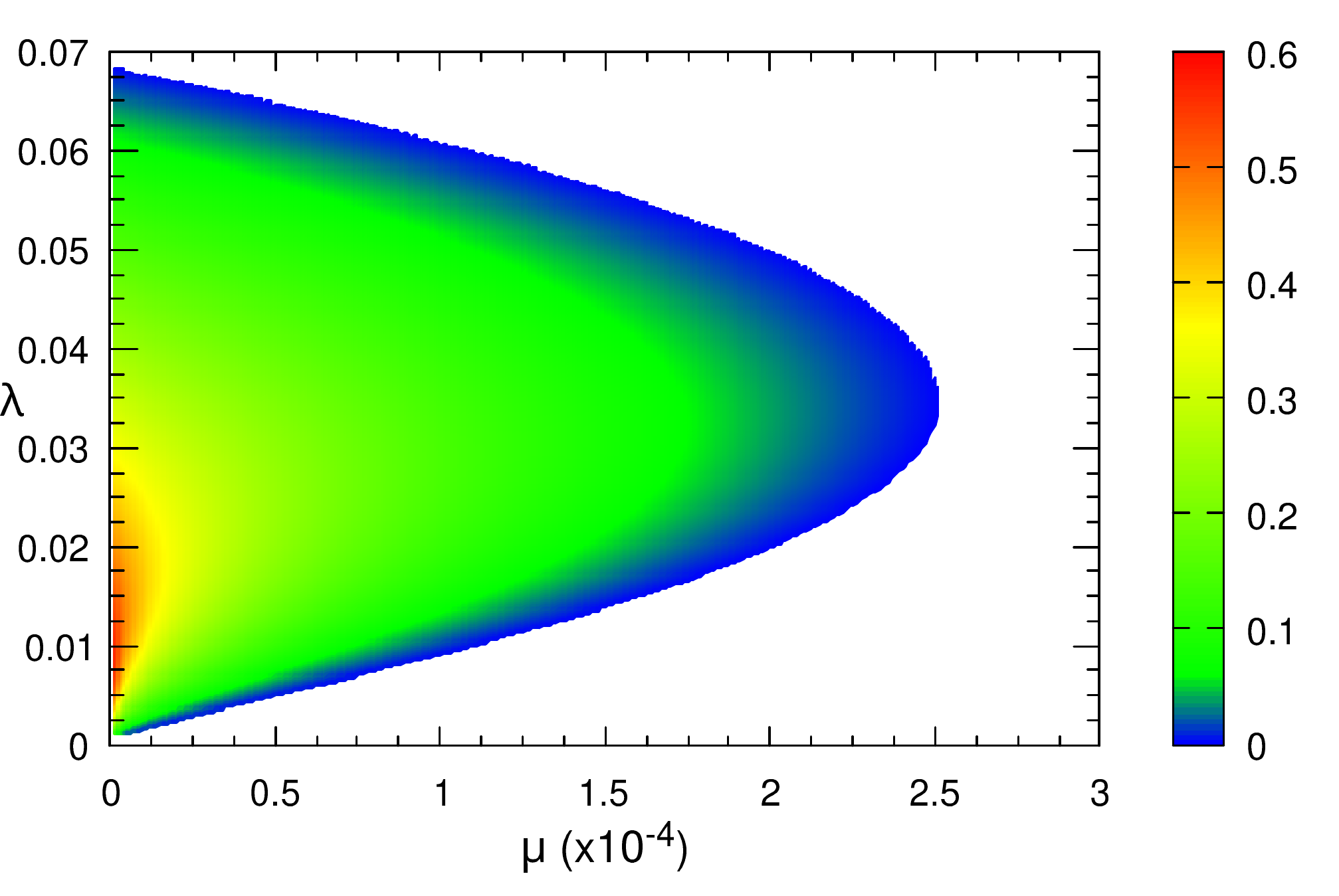}
}
\subfloat[]{
\includegraphics[angle=0,scale=0.4,clip]{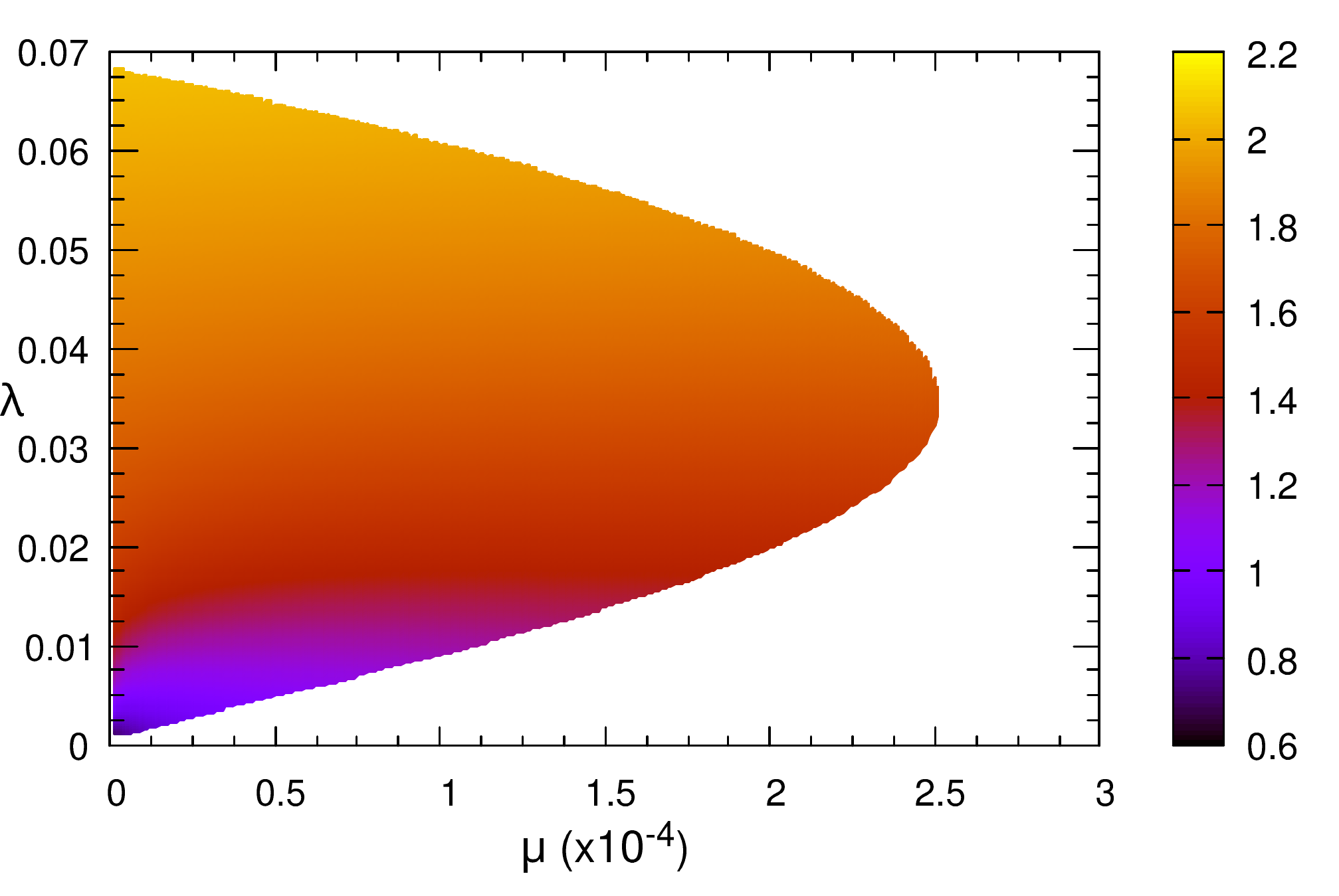}
}
\caption{\label{fig_Omega}
(a) Phase diagram in the plane {\tc $(\mu,\lambda)$}, depicting (a) the real part and (b) the imaginary part of the solution of \eref{eq_Omega_zeros2} for which the real part $\Re\, \Omega (>0)$ is the largest.}
\end{figure*}
%
In figure \ref{fig_surf2}, {\tc we have plotted} the modulus of \eref{eq_Omega_zeros2} in the complex $\Omega$-plane.
When $\lambda=0.08$ shown in (b), all zeros are located on the left part of the plane with negative real parts. For $\lambda=0.05$ in (a), only one conjugate pair of $\Omega$ has a strictly positive real part.
Figure \ref{fig_Omega} presents (a) the phase diagram in the plane {\tc $(\mu,\lambda)$} for
which there is at least one zero with a positive real part, together with (b) the corresponding frequency values for the rightmost zeros.
When the real part is positive ($\Re\, \Omega >0$), the perturbation is relevant and oscillations occur for any value of {\tc $(\mu,\lambda)$} in the colored domain except for $\mu=0$ where $\vk_1$ vanishes at some finite time.
{\tc As discussed previously,} for any nonzero value of $\mu$, $\vk_1$ does not vanish due to the presence of the special term $\mu/\vk_1$ in the integral of the right hand side of \eref{eq_v1_2}, and the oscillations are restricted to a range for which $\mu$ is no larger than $2.5\times 10^{-4}$. Outside the colored area of figure \ref{fig_Omega}(a), the perturbation is irrelevant and $\vk_1$ approaches its {\tc steady-state} value $\vk_1^*$.
{\tc We should also notice in figure \ref{fig_surf2} that if more than two conjugate pairs have positive real part, the spectrum of the oscillations should contain the different frequency components associated with each of these pairs.}

\begin{figure*}[ht]
\centering
\includegraphics[angle=0,scale=0.6,clip]{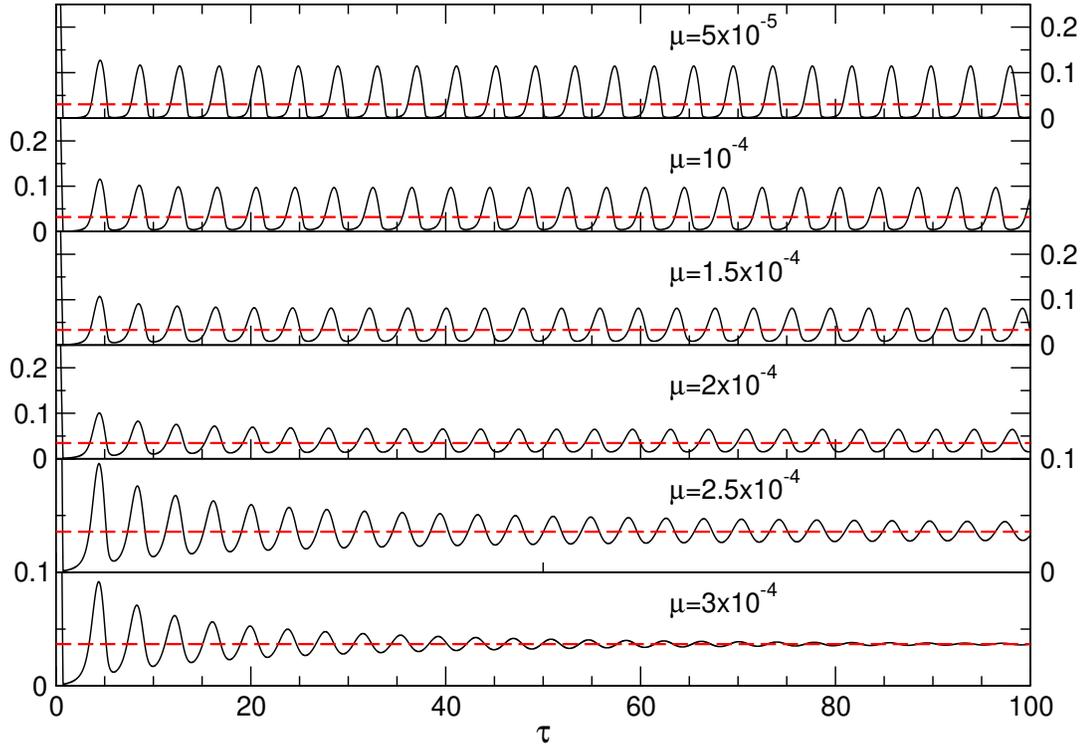}
\caption{\label{fig_osc2_N10000}
Time evolution of $\vk_1(\tau)$ for $N=10000$, $\lambda=0.03$, and several values of $\mu$. For $\mu=2.5\times 10^{-4}$, the oscillations are slowly damped and for $\mu=3\times
10^{-4}$ the oscillations disappear as $\vk_1$ approaches asymptotically $\vk_1^*$ (red dashed lines).}
\end{figure*}

{\tc Numerically, we have performed time resolution of \eref{eq_gen_vk} for $N=10000$ 
in figure \ref{fig_osc2_N10000} for a given $\lambda=0.03$ and for various values of $\mu$, using the RKF45 algorithm.
$\lambda$ was chosen such that it corresponds approximately to the maximum extension of the domain of stable oscillations in figure \ref{fig_Omega} for which oscillations up to $\mu=2.5\times 10^{-4}$ can be observed. For $\mu=3\times 10^{-4}$ oscillations are only transient since all zeroes have $\Re \Omega<0$. In Table \ref{tab:table2} we have analyzed the Fourier spectrum of 
the signals presented in figure \ref{fig_osc2_N10000} and reported their dominant frequency,
as well as the frequencies extracted from the solutions of \eref{eq_Omega_zeros2}. As expected the 
agreement between the two values is correct when the oscillation amplitudes are small
or, equivalently, near the onset of the oscillations.
}
%

%
\begin{table}[ht]
\centering
\caption{{\tc Frequency $\Omega_{\textrm{num}}$ for $\lambda=0.03$ and several values of $\mu$, extracted from the Fourier transform of the time oscillations in figure \ref{fig_osc2_N10000}. 
For comparison, the imaginary part $\Im\,\Omega$ of the rightmost zero of \eref{eq_Omega_zeros2} is also given.}}
\label{tab:table2}
{\renewcommand{\arraystretch}{1.2}
\begin{tabular}{c|c|c}
$\mu$ & $\Omega_{\textrm{num}}$ & $\Im\,\Omega$
\\  \hhline{===}
$5\times 10^{-5}$ & 1.548 &  1.684
\\
$1\times 10^{-4}$ & 1.573 & 1.655
\\
$1.5\times 10^{-4}$ & 1.594 & 1.640
\\
$2\times 10^{-4}$ & 1.612 & 1.632
\\
$2.5\times 10^{-4}$ & 1.627 & 1.628
\\
\hline
\end{tabular}}
\end{table}

\section{Discussion \label{sec_disc}}
%
We have shown that oscillations are present in both sets of
exponents $(\as,\bs)=(1,0)$ and $(1,1)$, but {\tc the maximum amplitude range of the parameters
$\mu$ and $\lambda$ for which these oscillations exist and are stable is quite different, even if exponent $\bs$ changes by only one unit. In the first case,
the domain of admissible parameters $(\mu,\lambda)$ extends approximately to the values $(0.75\times 10^{-5},1)$, whereas in the second case it is bounded by $(2.5\times 10^{-4},0.07)$, by inspection of figures \ref{fig_omega1_surf} and \ref{fig_Omega} respectively. The range of $\lambda$ is quite different, much smaller in the second model for which the maximum of $\mu$ is in contrast more than 10 times larger.} 
More generally, the oscillator equations (\ref{eq_v1}) and (\ref{eq_v1_2}) can be put into the generalized form:
\bb \label{eq_v1_gen}
\vk_1(\tau)=1-2
\int_0^{\tau} d\tau' \frac{\vk_1(\tau')\exp\left[
-\beta(\tau-\tau')-2\mu\int_{\tau'}^{\tau} d\tau'' \vk_1^{-1}(\tau'')\right]}{\left(
1-\int_{\tau'}^{\tau}d\tau'' \exp \left
[-\alpha(\tau-\tau'')-\mu\int_{\tau''}^{\tau}d\tau''' \vk_1^{-1}(\tau''')\right ]
\right )^3},
\ee
with the parameters
\bb\nn
(\as,\bs)=(1,0):\;\alpha\equiv 1 ~\mbox{and}~ \beta\equiv 2+\lambda,
\\
(\as,\bs)=(1,1):\;\alpha\equiv 1+\lambda ~\mbox{and}~ \beta\equiv 2+2\lambda .
\ee
{\tc The denominator in the integral is important for the existence of oscillating solutions. 
Indeed, let us consider the following class of less complex integral equations}
\bb \label{eq_v1_ex}
\vk_1(\tau)=1-2
\int_0^{\tau} d\tau' \vk_1(\tau')\exp\left[
-\beta(\tau-\tau')-2\mu\int_{\tau'}^{\tau} d\tau'' \vk_1^{-1}(\tau'')\right],
\ee
with $\beta>0$ for simplicity. From this integral equation, we obtain, via differentiation and reorganization of the different terms, a simple first-order ODE:
\bb \label{eq_v1_ex_diff}
\frac{d\vk_1(\tau)}{d\tau}=-(2+\beta)\vk_1(\tau)+\beta-2\mu+2\frac{\mu}{\vk_1(\tau)},
\ee
which has the positive asymptotic constant solution
\bb \label{eq_v1s_ex}
\vk_1^*=\frac{\beta-2\mu+\sqrt{(\beta+2\mu)^2+16\mu}}{2(2+\beta)} .
\ee
Again we can probe the stability of this solution by considering the exponential
perturbation $\vk_1(\tau)=\vk_1^*+\epsilon(\tau)$ with $\epsilon(\tau)=\epsilon_0e^{\Omega\tau}$.
A straightforward calculation leads to the unique real negative solution
\bb \label{omega_ex}
\Omega=-(2+\beta)-\frac{2\mu}{\vk_1^*}-\frac{4\mu}{\vk_1^*\beta+2\mu} < 0,
\ee
which implies that the class of integral equations (\ref{eq_v1_ex}) has no asymptotic solution other than the constant solution given by \eref{eq_v1s_ex}. It is easy to see that for $\mu=0$, the
solution of \eref{eq_v1_ex_diff}, which is given by $\vk_1(\tau)= (2+\beta)^{-1} (2e^{-(2+\beta)\tau}+\beta)$,
decays exponentially towards $\vk_1^*$ with rate $\Omega=-(2+\beta)<0$.

In general, one can examine the stability of \eref{eq_v1_gen} for any parameters $\alpha$ and $\beta$,
in order to generalize equations (\ref{eq_Omega_J1}) and (\ref{eq_Omega_J2}) for the complex frequency $\Omega$:
\bb\nn
1+2\fJ_3(\bar\beta{+}\Omega,\bar\alpha)+2\frac{\mu}{\Omega\vk_1^*}\Big(
2\Big [\fJ_3(\bar\beta,\bar\alpha)-\fJ_3(\bar\beta{+}\Omega,\bar\alpha)\Big ] 
\\ \label{eq_Omega_Jgen}
~~+\frac{3}{\bar\alpha}\Big [\fJ_4(\bar\beta,\bar\alpha)-\fJ_4(\bar\beta{+}\bar\alpha,\bar\alpha) \Big ]
-\frac{3}{\bar\alpha+\Omega}\Big [\fJ_4(\bar\beta,\bar\alpha)-\fJ_4(\bar\beta{+}\bar\alpha{+}\Omega,\bar \alpha)\Big ]
\Big )=0,
\ee
with $\bar\alpha \equiv \alpha+\mu/\vk_1^*$ and $\bar\beta \equiv \beta+2\mu/\vk_1^*$, where
$\vk_1^*$ is the {\tc long time} constant solution of \eref{eq_v1_gen} satisfying the implicit equation
\bb
{\vk_1^*}^{-1}=1+2\fJ_3(\bar\beta,\bar\alpha),
\ee
{\tc which is a generalization of equations (\ref{eq_v1eq2}) and (\ref{eq_v1_2lim})}.
For a general set of exponents $(\as,\bs)$, it is not obvious how to obtain a general oscillator equation
when $\as$ or $\bs$ can take fractional values. In this case, we can nevertheless generalize \eref{eq_phi1} for the moments $\su_{r}$ to any real number $r$:
\bb \label{eq_phi_gen}
\frac{\parti \su_r}{\parti\tau}=-\left(1+\frac{\mu}{\vk_1}\right)\su_{r+\as}-\lambda\su_{r+\bs}
+\sum_{j=0}^{\infty}{r+\as\choose j}\su_j+2^{r+\as}\vk_1 .
\ee
It is convenient to introduce the set of generating functions with positive
integers $(l,m,n)$:
\bb \label{eq_Gmn}
G_{mn}(u,\tau)\equiv\sum_{l=0}^{\infty}\frac{u^l}{l!}\su_{l+m\as+n\bs},
\ee
since \eref{eq_phi_gen} should generate via recursion all possible moments with $r=l+m\as+n\bs$,
which form a set of closed indices for this equation.
While initial conditions are given by $G_{mn}(u,0)=0$,
we are interested in the first component $G_{00}(0,\tau)=\su_0=1-\vk_1$.
Multiplying \eref{eq_phi_gen} by $u^l/l!$ and summing over $l$ with $r=l+m\as+n\bs$, we obtain
\bb\nn
\frac{\partial G_{mn}}{\partial\tau}&=&
-(1+\mu/\vk_1)G_{m+1n}-\lambda G_{mn+1}
+2^{(m+1)\as+n\bs}e^{2u}\vk_1
\\ \label{eq_Gmn_diff0}
& &+\sum_{l=0}^{\infty}\frac{u^l}{l!}\sum_{j=0}^{\infty}{l+(m+1)\as+n\bs\choose j}\su_j ,
\ee
where the double sum can be evaluated through the use of the Egorychev method as in \eref{eq_binom}:
\bb
\sum_{l\ge 0}^{}\frac{u^l}{l!}\sum_{j\ge 0}^{}{l+(m+1)\as+n\bs\choose j}\su_j
=\sum_{l\ge 0}\frac{u^l}{l!}\sum_{j\ge 0}
\oint \frac{dz}{2i\pi z}\frac{(1+z)^{l+(m+1)\as+n\bs}}{z^j}\su_j .
\ee
This yields
\bb\nn
\sum_{l\ge 0}^{}\frac{u^l}{l!}\sum_{j\ge 0}^{}{l+(m+1)\as+n\bs\choose j}\su_j
&=&\sum_{j\ge 0}\oint \frac{dz}{2i\pi z}\frac{(1+z)^{(m+1)\as+n\bs}}{z^j}e^{u(1+z)}\su_j
\\ \nn
&=&\sum_{l\ge 0}\sum_{j\ge l}e^u{(m+1)\as+n\bs\choose l}\frac{u^{j-l}}{(j-l)!}\su_j
\\
&=&e^u\sum_{l\ge 0}{(m+1)\as+n\bs\choose l}\frac{\partial^lG_{00}(u,\tau)}{\partial u^l},
\ee
which leads the closed form for \eref{eq_Gmn_diff0}:
\bb\nn
\frac{\partial G_{mn}}{\partial\tau}&=&
-(1+\mu/\vk_1)G_{m+1n}-\lambda G_{mn+1}
+2^{(m+1)\as+n\bs}e^{2u}\vk_1
\\ \label{eq_Gmn_diff}
& &+e^u\sum_{l\ge 0}{(m+1)\as+n\bs\choose l}\frac{\partial^l G_{00}}{\partial u^l}.
\ee
The last sum can formally be identified as the derivation operator $(1+\partial_u)^{\gamma}$
with $\gamma=(m+1)\as+n\bs$. In the previous analysis with $(\as,\bs)=(1,0)$ and $(\as,\bs)=(1,1)$, see equations (\ref{eq_G}) and (\ref{eq_G2}), we have, for $(m,n)=(0,0)$, $\gamma=1$, and therefore the
{\tc sum overl $l$ in \eref{eq_Gmn_diff} becomes simply $e^{u}(G_{00}+\partial_uG_{00})$}.
In the general case of $\as$ and $\bs$ integers, we obtain the following linear partial differential equation
with non-constant coefficients:
\bb\nn
\frac{\partial G_{00}}{\partial\tau}&=&
-(1+\mu/\vk_1)\frac{\partial^{\as} G_{00}}{\partial u^{\as}}-
\lambda \frac{\partial^{\bs} G_{00}}{\partial u^{\bs}}
+2^{\as}e^{2u}\vk_1
+e^u\left (1+\frac{\partial}{\partial u}\right)^{\as}G_{00},
\ee
which is in general unsolvable by means of characteristics curves, except for the cases treated in the previous sections.
%
\section{Conclusion \label{sec_concl}}
%
In this paper, we have investigated the dynamics of an infinite set of clusters interacting with monomers leading to aggregation or coagulation with rate $k^{\as}$ or fragmentation with rate $\lambda k^{\bs}$, depending on the mass $k$ of the cluster. This usually leads to an equilibrium state, but we have shown that the addition of a self-disintegration process with rate $\mu k^{\as}$ can induce oscillations in the cluster densities for a restricted domain of parameters {\tc $(\mu,\lambda)$}, where $\mu$ can be infinitely small but non-zero. The cluster densities $\vk_k$ are solely a function of the monomer density $\vk_1$, and the latter is
determined implicitly by a time integral equation in the form of \eref{eq_v1_gen}. This implicit equation always possesses a non-zero {\tc steady-state} solution. 

{\tc The stability of this constant solution can be probed by introducing a small time perturbation of exponential form with a complex rate $\Omega$ satisfying an implicit equation, see \eref{eq_Omega_Jgen}.
If the real part of all the solutions $\Omega$ is negative, the constant solution is stable. Otherwise, if a finite set of solutions has positive real parts, oscillations are spontaneously generated. Their frequency is related to the imaginary part of the most relevant solutions $\Omega$, e.g. with the largest
real part. The frequency of these oscillations is generally slightly different and
this difference is negligible on the boundary of the domain of oscillations when $\Re\,\Omega=0$,
and increases with $\Re\,\Omega >0$}. However, we cannot extract from the function (\ref{eq_Omega_Jgen}) the exact oscillation frequency since the integral \eref{eq_v1_gen} is highly nonlinear. Corrections to the estimated frequency can be made through the use of standard techniques \cite{book:Mook} that can be applied if we know the existence of an ODE for $\vk_1$, see \cite{Fortin:2022} in the case of the Li\'enard mechanism. However, there is no obvious ODE of finite order derivable from \eref{eq_v1_gen} through consecutive differentiation and recursive substitutions in a controllable manner.
We also notice that collective oscillations exist in finite systems even if their amplitude vanishes in the limit of the system size $N \rightarrow \infty$, see figures \ref{fig_oscN1000} and \ref{fig_oscN10000}. This is due to the variations with $N$ of the location of the solutions $\Omega$ of \eref{eq_Omega_Jgen} in the complex plane.

\ack
This work was supported by the National Research Foundation of Korea through the Basic Science Research Program (Grant No. 2022R1A2C1012532).

\section*{References}

\bibliography{biblio-cluster}
\end{document}